\documentclass[%
 reprint,
 amsmath,amssymb,
 aps,
]{revtex4-2}

\usepackage{amsmath}
\usepackage{physics}
\usepackage{graphicx}
\usepackage{dcolumn}
\usepackage{bm}

\begin{document}

\preprint{APS/123-QED}

\title{From localized to well-mixed:\\How commuter interactions shape disease spread}

\author{Aaron Winn}
\email{winna@sas.upenn.edu}
\author{Adam Konkol}
\author{Eleni Katifori}

\affiliation{%
 Department of Physics and Astronomy, University of Pennsylvania, Philadelphia, Pennsylvania 19104, USA
}%

\date{\today}

\begin{abstract}
Interactions between commuting individuals can lead to large-scale spreading of rumors, ideas, or disease, even though the commuters have no net displacement. The emergent dynamics depend crucially on the commuting distribution of a population, that is how the probability to travel to a destination decays with distance from home. Applying this idea to epidemics, we will demonstrate the qualitatively different infection dynamics emerging from populations with different commuting distributions. If the commuting distribution is exponentially localized, we recover a reaction-diffusion system and observe Fisher waves traveling at a speed proportional to the characteristic commuting distance. If the commuting distribution has a long tail, then no finite-velocity waves can form, but we show that, in some regimes, there is nontrivial spatial dependence that the well-mixed approximation neglects. We discuss how, in all cases, an initial dispersal-dominated regime can allow the disease to go undetected for a finite amount of time before exponential growth takes over. This “offset time” is a quantity of huge importance for epidemic surveillance and yet largely ignored in the literature. 
\end{abstract}

\maketitle


\section{\label{sec:level1}Introduction}

The spread of an infectious disease occurs on several different length scales  \cite{Balcan_Colizza_Goncalves_Hu_Ramasco_Vespignani_2009, Wang_Li_2014, Riley_2007}. Long-distance disease transmission has been modeled on global air transport networks by utilizing known air travel rates between major cities \cite{Rvachev_Longini_1985, Brockmann_Helbing_2013, Colizza_Barrat_Barthelemy_Vespignani_2006}. At smaller scales, disease can spread within a country along daily commuting networks, and spatial correlations in infection patterns can be observed \cite{Charaudeau_Pakdaman_Boelle_2014, Treut_Huber_Kamb_Kawagoe_McGeever_Miller_Pnini_Veytsman_Yllanes_2021, Jia_Lu_Yuan_Xu_Jia_Christakis_2020, Calvetti_Hoover_Rose_Somersalo_2021, Gatto_Bertuzzo_Mari_Miccoli_Carraro_Casagrandi_Rinaldo_2020}. In either case, the fundamental challenge is to determine when a disease will reach a particular location, as quantified by the infection arrival time \cite{Gautreau_Barrat_Barthelemy_2007}.  The spatial distribution of arrival times depends on the human mobility patterns. On large scales, the arrival time can be estimated by introducing an effective distance related to the frequency of flights between major cities \cite{Iannelli_Koher_Brockmann_Hovel_Sokolov_2017}. On smaller scales, the arrival time is largely determined by the geographic separation between locations. Long-distance travel can quickly turn a local outbreak into a global epidemic. However, once an infectious disease has established itself in a city, the small-scale commuter dynamics become the dominant source of new cases \cite{Russell_Wu_Clifford_Edmunds_Kucharski_Jit_2021}. The goal of this paper is to demonstrate how different commuting patterns give rise to different patterns of disease spread.

Early work studying the spatial spread of infectious disease consisted of continuous models in a uniform population treating the contact rate between individuals as nonlocal \cite{Kendall_1957, Kendall_1965}. When the contact distribution is sufficiently localized, a reaction-diffusion system is recovered, and far from the infection’s origin, waves are observed \cite{Mollison_1972a}. An overview of the early work on spatial epidemic models can be found in \cite{Rass_Radcliffe_2003}. These early papers thoroughly studied traveling waves and critical behavior of reaction-diffusion systems \cite{Diekmann_1978, Hosono_Ilyas_1995}, but neglected important aspects of a realistic epidemic including the spatially discrete nature of case count data, the asymmetry in the contact distribution, the effects of a nonuniform population, and the precise meaning of the contact distribution as it relates to human mobility. 

More recent metapopulation modeling has addressed these theoretical issues. In metapopulation modeling, populations are grouped together into patches where compartmental models can be defined. These patches are isolated, except for some coupling between them \cite{Martcheva_2015}. One study by Dirk Brockmann \cite{Brockmann_2010} connected human mobility patterns in a metapopulation model to the continuous reaction-diffusion systems studied decades ago. There, conservation of population in each patch was enforced by imposing detailed balance on the flux of travelers. Our study revisits some of the calculations performed by Brockmann in a model more appropriate for commuter dynamics where local population conservation is inherent, and focuses on a surprisingly ignored aspect of infectious spread: the behavior close to the infection origin. The fact that these reaction-diffusion systems fail to account for long-distance air travel suggests that these models are most applicable over length scales where commuter travel dominates. Not only is the behavior close to the origin theoretically interesting in that it deviates from the well-studied waves displayed at long distances, but it is also the most important regime to study if we care to apply the model to real geographic regions. For many cases, we will see that competition between human dispersal and exponential growth in the number of infections gives rise to a characteristic time at which the disease at any particular location is detected. 

The paper is structured as follows. First, we revisit the simple assumptions and consequences of the spatially independent $SIR$ model. Next, we introduce a general metapopulation model based on interactions between two traveling individuals. From this, we focus on localized travel and derive our system of reaction-diffusion equations. Features of these equations are studied numerically and analytically. Finally, we test whether these features survive when used with more complicated and realistic metapopulation models on maps. A reference of important definitions of functions and parameters is given in the appendix. 

\section{Constructing the model}
\subsection{$SIR$ model}\label{sec:SIR}
Before attempting to write down a space-dependent epidemic model, it is worth reviewing the assumptions that go into the widely used $SIR$ model \cite{Hethcote_2000}. Individuals capable of contracting a disease are labeled susceptible. Individuals currently sick and capable of infecting others are labeled infected. Individuals no longer sick are granted life-long immunity and are labeled recovered. The susceptible $S_{\text{tot}}(t)$, infected $I_{\text{tot}}(t)$, and recovered $R_{\text{tot}}(t)$ populations are tracked with time. Here we use the subscript ``tot'' to refer to the populations in the entire spatial domain, a distinction which will be necessary once we begin dividing our spatial domain into patches. There are only two parameters in the model: the recovery rate $\gamma$ and the contact rate $\beta$. The recovery rate is simply the inverse of the mean infectious period. The contact rate is a product of two quantities: the expected number of close contacts an individual makes per day (``close" in the sense that there is an opportunity for the disease to spread) and the probability that a close contact with an infected individual will infect a susceptible individual. All individuals in the population are treated equally in that the same rate constants are used for each individual. Furthermore, the population of size $N_{\text{tot}}$ is assumed to be well-mixed such that contact between any two individuals is equiprobable. 

In order to write down the model, one only needs to consider the change in infections in a time $\Delta t$. The product $(\beta \Delta t) \frac{I_{\text{tot}}}{N_{\text{tot}}}$ gives the probability that a susceptible individual will be infected in a time $\Delta t$. The total number of new infections must then be weighted by the current susceptible population. Additionally, $(\gamma \Delta t) I_{\text{tot}}$ individuals recover. One can then obtain the rate equations for the extensive quantities $S_{\text{tot}}$, $I_{\text{tot}}$, and $R_{\text{tot}}$, but it is often more convenient to work with the intensive quantities $s_{\text{tot}}\equiv S_{\text{tot}}/N_{\text{tot}}$, $i_{\text{tot}}\equiv I_{\text{tot}}/N_{\text{tot}}$, and $r_{\text{tot}}\equiv R_{\text{tot}}/N_{\text{tot}}$.
\begin{subequations} \label{eq:SIR}
    \begin{eqnarray}
    \frac{ds_{\text{tot}}}{dt} &=& -\beta s_{\text{tot}} i_{\text{tot}} \label{eq:Sir}\\
    \frac{di_{\text{tot}}}{dt} &=& +\beta s_{\text{tot}} i_{\text{tot}} -\gamma i_{\text{tot}} \label{eq:sIr}\\
    \frac{dr_{\text{tot}}}{dt} &=& +\gamma i_{\text{tot}}\label{eq:siR}
    \end{eqnarray}
\end{subequations}

An important property of these equations is that the total population is conserved. 
\begin{equation}
    \frac{d}{dt} (s_{\text{tot}}+i_{\text{tot}}+r_{\text{tot}}) = 0
\end{equation}
We can rescale time to eliminate one of the $\beta$, $\gamma$ rates leaving only a single dimensionless parameter $R_0 \equiv \beta/\gamma$ governing the infection dynamics. This is the basic reproduction number which can be interpreted as the expected number of people an infected person infects in a completely susceptible population. If $R_0<1$, the disease-free equilibrium is stable, leading to exponential decay of infections. If $R_0>1$, the disease-free equilibrium is unstable, leading to exponential growth in infections until herd immunity is reached, at which point $i_{\text{tot}}(t)$ reaches a peak value 
\begin{equation}
    i_{\text{max}} = 1 - \frac{1+\log (R_0 )}{R_0}. \label{eq:imax}
\end{equation}
and begins to fall to zero. The fraction of individuals who are at any point infected is given by solving the following transcendental equation for $r_{\text{tot}}(\infty)$:
\begin{equation}
    r_{\text{tot}}(\infty)+\frac{\log (1-r_{\text{tot}}(\infty))}{R_0} =0. \label{eq:rinf}
\end{equation}
The signatures $i_{\text{max}}$ and $r_{\text{tot}}(\infty)$ will even be present in the space-dependent models we will derive later.

One final quantity which will have an important analogy in the space-dependent model is the arrival time $t_{\text{arrival}}$, defined as the time at which $i_{\text{tot}}$ exceeds some small threshold value $i_c$. At early times when $s_{\text{tot}} \approx 1$, the linearization of \eqref{eq:sIr} gives
\begin{equation}
    i_{\text{tot}}(t) \approx i_{\text{tot}}(0) e^{(\beta - \gamma )t}.\label{eq:linearizedSIR}
\end{equation}
Equating this expression to $i_c$ and solving for $t$ gives 
\begin{equation}
    t_{\text{arrival}} = \frac{1}{(\beta - \gamma)} \log \left(\frac{i_c}{i_{\text{tot}}(0)}\right). \label{eq:tArrTot}
\end{equation}

\subsection{Metapopulation epidemic models}
Spatial epidemic models were originally formulated for a uniform-density continuous population by Kendall \cite{Kendall_1957}. He argued that the condition for criticality and the final epidemic size are identical to that of the spatially independent model. Later, he showed that by approximating the interactions between susceptible and infected individuals by a term of the form $s\nabla^2 i$, one finds that the disease spreads as a wave with a minimum speed \cite{Kendall_1965}, analogous to the result by Fisher \cite{Fisher_1937}. Mollison proved the existence of waves with a minimum speed without having to resort to the diffusion approximation by carefully choosing an exponential form for the contact rates \cite{Mollison_1972b}. He also showed that if the contact distribution is at least exponentially localized, then wave solutions can be expected, but otherwise, the infection propagates at an infinite speed \cite{Mollison_1972a}. 

More recent works define epidemic models in metapopulations. Metapopulation epidemic models can further be divided into two categories depending on how human mobility is incorporated \cite{Cosner_Beier_Cantrell_Impoinvil_Kapitanski_Potts_Troyo_Ruan_2009, Martcheva_2015}. In Lagrangian movement models, individuals have a well-defined home patch, but may make contact with individuals outside their home patch, spreading disease. The number of infected individuals who reside in a particular location is tracked with time. In Eulerian models, individuals migrate between patches. The number of infected individuals currently occupying a given patch is tracked with time. The terms ``Eulerian" and ``Lagrangian" parallel their use in fluid dynamics. For Lagrangian models, the dynamics track the people while for Eulerian models, the dynamics track the patch. The predictions made by these two classes of mobility models are not the same \cite{Cosner_Beier_Cantrell_Impoinvil_Kapitanski_Potts_Troyo_Ruan_2009}. Each of these models possesses advantages and disadvantages. Lagrangian models automatically satisfy the intuitive constraint that the total population of each patch is constant in time, reflecting the cyclic nature of commuting patterns. Lagrangian models only keep track of human mobility in as much as it is necessary to define nonlocal contact rates. This may seem like a simpler way to capture the propagation of disease without focusing too much on the motion of individuals, but relating the nonlocal contact rates to human mobility is more subtle in the Lagrangian formulation. This has led to some variation in the formulation of Lagrangian models \cite{Calvetti_Hoover_Rose_Somersalo_2021, Yin_Wang_Xia_Dehmer_Emmert-Streib_Jin_2020, Charaudeau_Pakdaman_Boelle_2014, Treut_Huber_Kamb_Kawagoe_McGeever_Miller_Pnini_Veytsman_Yllanes_2021}. 

The Eulerian models do not inherently satisfy population conservation unless detailed balance is assumed \cite{Brockmann_2010}. At small geographic scales, this seems like an unnatural assumption since it allows for the possibility of two individuals to swap positions permanently, but it has been found to be a good approximation on the worldwide air-transportation network \cite{Barrat_Barthelemy_Pastor-Satorras_Vespignani_2004}. The master equation describing human mobility can be easily written down in an Eulerian model. Rvachev and Longini first presented an Eulerian model \cite{Rvachev_Longini_1985}, and later work focused on calculating critical behavior \cite{Arino_Driessche_2005} and effective network distances governing the arrival time \cite{Gautreau_Barrat_Barthelemy_2007,Iannelli_Koher_Brockmann_Hovel_Sokolov_2017}. Some models have combined aspects of these models by taking into account explicit Eulerian movement of individuals while also enforcing that the individuals travel back to a well-defined home, as is typically only done in Lagrangian models \cite{Belik_Geisel_Brockmann_2011, Lipshtat_Alimi_Ben-Horin_2021, Arino_Driessche_2005, Arino_van_den_Driessche_2003}. 

\subsection{Lagrangian movement model}
Before introducing our commuter model, we present a mathematical introduction to epidemic models incorporating Lagrangian movement of individuals. In (Lagrangian) metapopulation models, we replace each of our differential equations \eqref{eq:Sir}, \eqref{eq:sIr}, \eqref{eq:siR} with $n_C$ differential equations, one for each of the $n_C$ patches, which will also be referred to as counties in the rest of this work. The contact rate $\beta$ is replaced by a contact matrix $\boldsymbol{\beta}$ with elements $\beta_{nk}$ describing the interaction between susceptible individuals who live at $n$ and infected ones who live at $k$. In continuous models, $\boldsymbol{\beta}$ is also referred to as the contact distribution. A precise physical interpretation of $\boldsymbol{\beta}$ will be made clear in the next section. 

The space-dependent rate equations now read:
\begin{subequations}
    \label{eq:SIRL}
    \begin{eqnarray}
        \frac{ds_n}{dt} &=& - s_n\sum_k \beta_{nk} i_k \label{eq:SirL}\\
        \frac{di_n}{dt} &=& + s_n \sum_k \beta_{nk} i_k -\gamma i_n \label{eq:sIrL}\\
        \frac{dr_n}{dt} &=& +\gamma i_n\label{eq:siRL}.
    \end{eqnarray}
\end{subequations}

The symbol $i_n$ denotes the proportion of infected people at $n$, $i_n = I_n/N_n$, where $I_n$ is the number of people living at $n$ who are infected, and $N_n$ is the total number of people living at $n$. Keeping in the spirit of the $SIR$ model, the total number of contacts an individual makes is the same for everyone; that is, we can define the $n$-independent quantity
\begin{equation}
    \beta \equiv \sum_k \beta_{nk}.
\end{equation}
This system also possesses a disease-free equilibrium when $s_n = 1, i_n = 0, r_n=0$ for all $n$.
The stability of the disease-free equilibrium will be determined by the eigenvalues of the Jacobian
\begin{equation}
    J_{nk} \equiv \frac{\partial i_n}{\partial i_k}\Bigg|_{s_n=1} = \beta_{nk}-\gamma \delta_{nk}
\end{equation}
of the infected compartment.
The vector $(1,..., 1)^T$ is always an eigenvector of this linearized system with eigenvalue $\beta - \gamma$, so if $\beta-\gamma > 0$, the disease-free equilibrium is unstable. The converse is also true: if $\beta-\gamma < 0$, the disease-free equilibrium is stable. To show this, we need to demonstrate that in this case all eigenvalues of $\mathbf{J}$ have negative real parts. The following argument is adapted from \cite{Post_DeAngelis_Travis_1983}. Since $-\mathbf{J}$ is a $Z$-matrix (nonpositive off-diagonal entries), $-\mathbf{J}$ has eigenvalues with positive real parts if and only if there exists a $\vec{v}>0$ such that  $-\mathbf{J}\vec{v} > 0$ (in which case $-\mathbf{J}$ is an $M$-matrix). Using $\vec{v} = (1,...,1)^T$, we have $-\mathbf{J}\vec{v} = -(\beta - \gamma)\vec{v}$, so, if $\beta-\gamma < 0$, then the eigenvalues of $\mathbf{J}$ have negative real parts.  Thus, we can define  $R_0\equiv \beta /\gamma$, and the condition for stability is identical to that of the spatially independent model.

We can estimate the final size of the epidemic at each location $r_n(\infty)$ by multiplying equation \eqref{eq:siRL} by $\beta_{mn}$, summing over $n$, and dividing by equation \eqref{eq:SirL}:
\begin{equation}
    \frac{d\bar{r}_n}{ds_n} = -\frac{1}{R_0 s_n} \implies \bar{r}_n(\infty) + \frac{\log(1-\bar{r}_n(\infty))}{R_0} =0
\end{equation}
where $\bar{r}_n \equiv \sum_k \frac{\beta_{nk}}{\beta} r_k$. Notice, $\bar{r}_n(\infty)$ satisfies the same equation as $r_{\text{tot}}(\infty)$, and is therefore spatially independent. The solution to the linear equation relating $\bar{r}_n(\infty)$ to $r_k(\infty)$ is to have $r_k(\infty)=\bar{r}_n(\infty)=r_{\text{tot}}(\infty)$. The contact matrix encodes important spatial information, but the final size of the epidemic is still fixed by $R_0$.

\subsection{Commuter model}
The previous subsection establishes the general framework for Lagrangian metapopulation models in terms of the contact matrix $\boldsymbol{\beta}$. Next, we will derive a particular form of  $\boldsymbol{\beta}$ used to describe commuter dynamics.  The method of derivation will be similar to that in section \ref{sec:SIR}, only now, we must find $\Delta I_n$, the increase in infections at county $n$ in a time $\Delta t$. We imagine that an individual travels to work, makes contact with people from other counties, and then returns home. Recent work by Le Treut, et al. has demonstrated how a data-driven Lagrangian formalism can predict the spread of Covid-19 \cite{Treut_Huber_Kamb_Kawagoe_McGeever_Miller_Pnini_Veytsman_Yllanes_2021}. Some theoretical work has focused on identifying emergent length scales characterizing the spread of disease on a network \cite{Brockmann_Helbing_2013,Iannelli_Koher_Brockmann_Hovel_Sokolov_2017}. One work by Dirk Brockmann was particularly influential to our study in attempting to connect the travel rates in an Eulerian framework to the diffusion of infection in real space \cite{Brockmann_2010}. Our Lagrangian approach follows directly from a Bayesian derivation and considers the most general interactions where both susceptible and infected individuals travel. Many of the results will parallel those derived by Brockmann, but it is important to re-assert that the quantities we consider in the Lagrangian formalism are not identical to the related quantities in the Eulerian formalism. 

\begin{figure*}
    \centering
    \includegraphics[width=.95\textwidth]{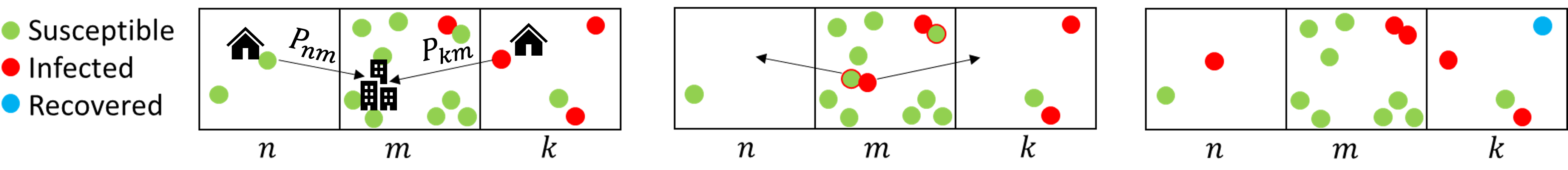}
    \caption{Commuter dynamics. A susceptible individual from county $n$ and an infected individual from $k$ make contact at $m$ and return home. A green circle with a red boundary indicates an infectious contact event.}
    \label{fig:commuters}
\end{figure*}

Let $P_{nm}$ denote the fraction of contacts made by the population living at location $n$ that occur at location $m$, with $\sum_m P_{nm}=1$. The  quantity $P_{nn}$ denotes the fraction of contacts made in the county of residence. Roughly speaking, $P_{nm}$ can be thought of as the probability of an individual who lives at $n$ traveling to $m$, so we will simply refer to $P_{nm}$ as the \textit{commuting probability}. This probability could depend on the separation of the counties as well as on the populations of any county, but we assume that $P_{nm}$ is independent of disease status. When a person travels from $n$ to $m$, they will encounter not only individuals who live at $m$, but also individuals who live at other locations $k$ that have traveled to county $m$. An example interaction between a susceptible and an infected individual is shown in figure \ref{fig:commuters}. The probability to make infectious contact at $m$ is equal to the fraction of people currently at $m$ (in the Eulerian sense) who are infected, regardless of where our traveler lives. With this in mind, we can calculate the probability that a worker from $n$ makes infectious contact.
\begin{align}
    \text{Pr}(\text{inf. contact}|\text{live at }n) 
    &= \sum_m \text{Pr}(\text{travel to }m|\text{live at }n)\nonumber \\ &\hspace{.1cm}\times \text{Pr}(\text{inf. contact}|\text{travel to }m)\nonumber \\
    &= \sum_m P_{nm} \frac{\sum_k P_{km} I_k}{\sum_k P_{km}N_k} \label{eq:Pr(IC)}
\end{align}

This quantity plays the role of $I/N$ in the $SIR$ model. It gives the probability for a particular close contact made by someone living at $n$ to be infectious. We can now easily write down the full \textit{commuter model}: 
\begin{subequations} \label{eq:CommuterModelSIR}
\begin{eqnarray}
    \frac{dS_n}{dt} &=& - \beta S_n \sum_m P_{nm} \frac{\sum_k P_{km} I_k}{\sum_k P_{km}N_k} \label{eq:S} \\
    \frac{dI_n}{dt} &=& + \beta S_n \sum_m P_{nm} \frac{\sum_k P_{km} I_k}{\sum_k P_{km}N_k}  - \gamma I_n \label{eq:I}\\
    \frac{dR_n}{dt} &=& + \gamma I_n. \label{eq:R}
\end{eqnarray}
\end{subequations}
In terms of the intensive quantities, the model reads:
\begin{subequations} \label{eq:CommuterModelsir}
\begin{eqnarray}
    \frac{ds_n}{dt} &=& - \beta s_n \sum_m P_{nm} \frac{\sum_k P_{km} N_k i_k}{\sum_k P_{km}N_k}. \label{eq:s} \\
    \frac{di_n}{dt} &=& + \beta s_n \sum_m P_{nm} \frac{\sum_k P_{km} N_k i_k}{\sum_k P_{km}N_k}  - \gamma i_n \label{eq:i}\\
    \frac{dr_n}{dt} &=& + \gamma i_n.\label{eq:r}
\end{eqnarray}
\end{subequations}
All sums are taken over the $n_c$ counties. Similar equations have been derived by other authors \cite{Calvetti_Hoover_Rose_Somersalo_2021, Charaudeau_Pakdaman_Boelle_2014}. 

We can now identify
\begin{equation}
    \beta_{nk} = \beta \sum_{m} P_{nm} \frac{P_{km} N_k}{\sum_l P_{lm}N_l}, \label{eq:contactMtxPnm}
\end{equation}
where the fraction represents the probability that an encounter at $m$ is with someone from $k$. Notice that the contact matrix 
 $\beta_{nk}$ is not symmetric in a heterogeneous population. However, the matrix $N_n \beta_{nk}$ is precisely the number of close contacts made per time where one individual is from $n$ and one individual is from $k$; this is necessarily a symmetric matrix. In the special case of a homogeneous population, $P_{lm}=P_{ml}$, so $\beta_{nk}=(\mathbf{PP}^T)_{nk} = \beta_{kn}$. Here, it is particularly clear that we have one factor of $\mathbf{P}$ for the commutes of susceptible individuals and one for the commutes of the infected. We need both of these matrices in order to fully capture the pair interactions between susceptible and infected individuals. 

Although we assume $\beta$ is constant throughout the paper, it is briefly worth mentioning how one could modify the model to account for spatially varying contact rates. Because the contacts occur at the destination $m$, one would have to make the replacement $\beta \sum_m \rightarrow \sum_m \beta_m $. This causes the infection growth rate, $i_{n,\text{max}}$, $r_n(\infty)$, and properties related to the spreading to all have a strong space dependence. Since we are mainly interested in the spatial aspects of commuting patterns, we will avoid the complexity introduced by non-uniform contact rates.

In order to better understand the form of these equations and the relation to Eulerian models, let us see what happens when travel outside of the home county is small. Letting $P_{nm} = \delta_{nm} + \epsilon_{nm}$ and keeping only terms linear in $\epsilon$ leads to the following simplification:
\begin{align}
    &\beta S_n \sum_m P_{nm} \frac{\sum_k P_{km} I_k}{\sum_k P_{km}N_k}  \approx \beta S_n \frac{I_n}{N_n} \nonumber\\ &+ \beta \frac{S_n}{N_n}\sum_{m \neq n} \left(N_n P_{nm} + N_m  P_{mn} \right) \left(\frac{I_m}{N_m} - \frac{I_n}{N_n} \right). \label{eq:PR(IC)simplified}
\end{align}
This resembles an Eulerian $SIR$ model with symmetric travel flux $\beta(N_n P_{nm}+N_m P_{mn})$ \cite{Iannelli_Koher_Brockmann_Hovel_Sokolov_2017}. The first term describes local growth, and the terms in the sum describe movement of the disease. There are two dominant ways the disease can move: a susceptible travels from $n$ and meets an infected at $m$, or an infected travels from $m$ and meets a susceptible at $n$. The more general form \eqref{eq:Pr(IC)} also allows for both the susceptible and the infected to travel, as in figure \ref{fig:commuters}, which we will refer to as second-order processes (in $\mathbf{P}$). Interestingly, the second-order processes can be unimportant in predicting the spread of the infection wave for several choices of $\mathbf{P}$, but are crucial for understanding the spread in some cases. This simplified form is identical to that used by Le Treut, et al. \cite{Treut_Huber_Kamb_Kawagoe_McGeever_Miller_Pnini_Veytsman_Yllanes_2021} who neglect second-order processes. 

For numerical simulations, the more general commuter dynamics \eqref{eq:Pr(IC)} will always be used. Equation \eqref{eq:CommuterModelsir} will be solved numerically in Python using the solve\_ivp function in the SciPy library with method ``RK45".

\subsection{Human Mobility Theory}
While some metapopulation models incorporate known, specific travel rates for particular locations \cite{Treut_Huber_Kamb_Kawagoe_McGeever_Miller_Pnini_Veytsman_Yllanes_2021, Jia_Lu_Yuan_Xu_Jia_Christakis_2020, Charaudeau_Pakdaman_Boelle_2014, Calvetti_Hoover_Rose_Somersalo_2021, Brockmann_Helbing_2013}, we are interested to understand the basic principles underlying infection spread so we will take advantage of the research on general human mobility patterns. Traditionally, human mobility was described using gravity models, that is, models which assume that the flux of individuals traveling between two locations is proportional to the populations of the two counties multiplied by some decreasing function of the separation distance (such as a power law, truncated power law, exponential, or Gaussian). In such models, there exists at least one length scale describing how far people travel. See \cite{Barbosa_2018} for a review on models for human mobility. Recently, people have become interested in the radiation model of human mobility that describes an individual's tendency to search for work at the closest location a job is available \cite{Simini_Gonzalez_Maritan_Barabasi_2012}. This model does not have any inherent travel length scale, only emergent length scales derived from the population distribution. Although the radiation model could be incorporated into \eqref{eq:Pr(IC)}, the existence of an inherent length scale is going to make it easier to study the general features of our theory, so we will only focus on gravity models in this paper. Thus, we assume the travel probabilities can be cast in the form 
\begin{equation}
    P_{nm} = \frac{N_m^\zeta G( |\vec{x}_n - \vec{x}_m| )}{\sum_m N_m^\zeta G( |\vec{x}_n - \vec{x}_m| )}. \label{eq:mobility}
\end{equation}
where $\int G(|\vec{x}|) d^d x =1$. The function $G$ will be referred to as the (homogeneous) \textbf{commuting distribution}, since $G(|\vec{x}_m-\vec{x}_n|)$ describes how far people are willing to commute in a homogeneous population. Physically, the sum in the normalization is always taken over a domain containing a finite number of counties, so there is no need to worry about convergence. Still, it will be helpful to think about length scales in the problem that may, in some cases, allow us to take the continuum limit, and in special cases, allow us to derive local transport operators. 

Many works find that a truncated power law
\begin{equation}
    G(|\vec{x}|) = \frac{e^{-|\vec{x}|/l}}{(|\vec{x}|+r_0)^p}\Bigg{/}\int_\Omega \frac{e^{-|\vec{x}|/l}}{(|\vec{x}|+r_0)^p}d^d x\label{eq:TruncatedPowerLaw}
\end{equation}
with appropriate choices of $p, l, $ and $r_0\ll l$ describes human mobility, as summarized in \cite{Gallotti_Bazzani_Rambaldi_Barthelemy_2016}. Note that the special case of a power law can be obtained by setting $l=\infty$, and the special case of an exponential distribution corresponds to setting $p=0$. When both $p=0$ and $l=\infty$, the commuting distribution is uniform, so the population is well-mixed.

To demonstrate how to perform these calculations, first consider a uniform population on a square grid with lattice constant $\Delta x$ in $d$ dimensions. The domain will be denoted $\Omega$. Suppose the population changes slowly over a length scale greater than that of $\Delta x$. Then, we can take the continuum limit $\Delta x \rightarrow 0$ and work in terms of a continuous population density $N_m \rightarrow N(\vec{x}_m) d^dx_m $ as well as the probability density function
\begin{align}
    P_{nm} \rightarrow P(\vec{x}_n,\vec{x}_m)d^d x_m &= \frac{N(\vec{x}_m)^\zeta G(|\vec{x}_m-\vec{x}_n|)d^d x_m}{\int_\Omega  N(\vec{x}_m)^\zeta G(|\vec{x}_m-\vec{x}_n|) d^d x_m}. \label{eq:P(x,y)}
\end{align}
In many cases, $G(|\vec{x}_m-\vec{x}_n|)$ is presumably localized near $\vec{x}_m = \vec{x}_n$, so one could try to Taylor expand the factor of $N(\vec{x}_m)^\zeta$ around $\vec{x}_n$. However, if $G(|\vec{x}_m-\vec{x}_n|)$ is a power law, then some of the moments of $\vec{x}$ will diverge. Mollison proved that wave solutions exist if and only if the contact distribution falls off faster than exponential at long distances \cite{Mollison_1972a}. For our problem, this suggests that we can try to replace our integral expressions with local differential operators (acting on $N$ and $i$) only if $G(|\vec{x}|)$ falls off faster than exponential. In this case, we can approximate
\begin{equation}
    P(\vec{x}_n,\vec{x}_m) \approx \left[1 + \zeta \partial_\mu \log N(\vec{x}_n) (x_m^\mu - x_n^\mu) \right] G(|\vec{x}_m-\vec{x}_n|)  \label{eq:approxPnm}.
\end{equation}
This is a simpler form to work with, which makes it clear that there is a tendency that increases with $\zeta$ to travel toward higher populations. There were multiple approximations which went into putting the population in this simple form. We can summarize these approximations by comparing length scales. Let $l$ denote the characteristic length scale contained in $G$ describing how far people are willing to travel for work, and let $L$ denote the size of the domain.

\begin{align}
    \Delta x \ll l \hspace{2.4cm}\text{(continuum limit)} \label{eq:continuumLimit}\\
    x \ll L \hspace{1cm} \text{(neglect boundary effects)} \label{eq:x<L}\\
    \exists k>0 \text{ s.t.}  \int_{\Omega} e^{k|\vec{x}|} G(|\vec{x}|) d^d x < \infty \hspace{.75cm} \text{(localized)} \label{eq:Mollison}
\end{align}
Having $\Delta x$ smaller than other relevant length scales allows us to take the continuum limit. When our model is applied to data, $\Delta x$ can be interpreted as the county size, and it may not be the smallest length scale, but for studying our theory, there is nothing preventing us from refining the domain until this inequality is satisfied. When $L$ is much larger than other relevant length scales, we can ignore boundary effects allowing us to drop odd moments of $x$ and assert that the even moments are independent of $n$. Rewriting our integral equations in terms of local operators and moments is valid whenever \eqref{eq:Mollison} is satisfied. This technique is only helpful if we can truncate the Taylor series after a few terms, so one might think that $l$ needs to be less than the length scale over which the population changes in order to derive \eqref{eq:approxPnm}, but we will see that the short-wavelength population fluctuations will not change the large-scale infection dynamics. 

Localized commuting does not necessarily imply that the infection is confined to a small region of space, especially in small systems where the infection quickly reaches the boundary. To avoid confusion, the term \textit{localized} will be reserved for commuting distributions satisfying \eqref{eq:Mollison}, and commuting distributions such as power laws which do not satisfy \eqref{eq:Mollison} will be called \textit{delocalized}. Infectious spread which is confined to certain regions of space at a given time will be called \textit{heterogeneous}, and infection growth which is nearly uniform in space will be called \textit{homogeneous}. The term \textit{well-mixed} may be applied either to a population with spatially independent contact distribution $G(|\Vec{x}|)=1/L^d$ such that \eqref{eq:SIR} is recovered, or more generally to any contact processes which give rise to nearly homogeneous spreading, even if $G$ is non-constant. Finally, the term \textit{uniform population} refers to a population with $N_n=$ constant such that $\mathbf{P}$ and $\boldsymbol{\beta}$ are symmetric. 

\section{Continuum Results}
In the next two sections, we study the results of the commuter model. In this section, we will work under approximation \eqref{eq:continuumLimit} and assume the commuting probabilities take the form \eqref{eq:P(x,y)}. We will work in terms of the continuous infection density $I(\vec{x},t)$ and population density $N(\vec{x})$ by making the replacements
\begin{equation}
    I_n(t) \rightarrow I(\vec{x}_n,t)d^d x_n \label{eq:discreteI}
\end{equation}
\begin{equation}
    N_n \rightarrow N(\vec{x}_n)d^d x_n \label{eq:discreteN}
\end{equation}
\begin{equation}
    i(\vec{x},t) \equiv \frac{I(\vec{x})}{N(\vec{x})}.
\end{equation}
Note that, unlike $I(\vec{x},t)$ and $N(\vec{x})$, $i(\vec{x},t)$ is unitless. Simulations will always necessarily be performed using the discretized quantities, but when discretized on a square lattice with side length $\Delta x$, one can convert to the continuous infection density at the lattice site by dividing by $(\Delta x)^d$; for example, $I(\vec{x}_n,t) = I_n(t)/(\Delta x)^d$. Because the step size falls out of our continuous results, even when the underlying map over which $I_n$ is calculated is not a lattice, the continuum model will give us an estimate for $I_n$, provided that \eqref{eq:continuumLimit} is satisfied.

Analogous to the contact matrix \eqref{eq:contactMtxPnm}, we can define the contact distribution 
\begin{equation}
    \beta(\Vec{x},\Vec{z}) = \beta \int_\Omega d^d y \hspace{.1cm}P(\vec{x},\vec{y}) \frac{P(\vec{z},\vec{y}) N(\vec{z})}{\int_\Omega d^d w \hspace{.1cm}P(\vec{w},\vec{y})  N(\vec{w})}
\end{equation}
describing interactions between susceptible individuals at $\Vec{x}$ and infected individuals at $\Vec{z}$. Equation \eqref{eq:i} becomes
\begin{equation}
    \frac{\partial i(\vec{x},t)}{\partial t} =  s(\vec{x},t) \int_\Omega d^d z \hspace{.1cm} \beta(\Vec{x},\Vec{z}) i(\Vec{z},t) - \gamma i(\vec{x},t) \label{eq:iC}
\end{equation}
along with obvious continuum versions for equations \eqref{eq:s} and \eqref{eq:r}. 

We will encounter moments of the form 
\begin{equation}
    \langle x^\mu \rangle \equiv \int_\Omega (y^\mu - x^\mu) G(|\vec{y}-\vec{x}|) d^dy
\end{equation}
\begin{equation}
    \langle x^\mu x^\nu \rangle \equiv \int_\Omega  (y^\mu - x^\mu)(y^\nu - x^\nu) G(|\vec{y}-\vec{x}|)d^d y
\end{equation}
and so on. When $\Omega$ is bounded, all moments will be finite regardless of $G$, but in an infinite space, all moments will be finite if and only if \eqref{eq:Mollison} is satisfied. It is tempting to set all odd moments to zero and claim that all even moments are independent of home location $\vec{x}$, but this symmetry is only exact in an infinite space or at the origin $\vec{x}=0$ and will fail considerably near the boundaries. Because no individuals can travel outside of $\Omega$, we must have that the infection flux through the boundary is zero. This amounts to imposing Neumann boundary conditions
\begin{equation} \label{eq:Neumann}
    \grad i(\vec{x},t) \cdot \hat{n} = 0, \hspace{.5cm} \vec{x} \in \partial \Omega.
\end{equation}

\subsection{The diffusion approximation for localized commutes}
To begin our analysis of the continuous commuter model, we assume that \eqref{eq:Mollison} is satisfied such that we can perform Taylor expansions like those performed to arrive at equation \eqref{eq:approxPnm}. Furthermore, we will assume that \eqref{eq:x<L} is satisfied. We will still, in general, consider the problem on a finite domain with \eqref{eq:Neumann} satisfied, but \eqref{eq:x<L} will allow us to simplify the moments as follows:
\begin{equation}
    \langle x^\mu \rangle \approx 0
\end{equation}
\begin{equation}
    \langle x^\mu x^\nu \rangle \approx \frac{1}{d}\int |\vec{y}-\vec{x}|^2 G(|\vec{y}-\vec{x}|) d^d y \hspace{.1cm} \delta^{\mu \nu}  \equiv \frac{1}{d} \langle x^2 \rangle \delta^{\mu \nu}.
\end{equation}
That is, even though our space is finite, we will be using the moments derived from an infinite space where all of the odd moments vanish and the even moments are independent of the home county. 

To simplify \eqref{eq:iC}, we will take advantage of the localized nature of $G(|\vec{x}-\vec{y}|)$ and perform a series of Taylor expansions up to second order in derivatives of $N$ and $i$. The details are shown in Appendix A, but the resulting equation is
\begin{equation}
    \frac{\partial i}{\partial t} = (\beta s-\gamma) i +  \frac{\beta \langle x^2 \rangle}{d} s \left[ \nabla^2 i + \grad \log N \cdot \grad i \right]. \label{eq:reactionDiffusion}
\end{equation}
The only mention of our original commuting probability shows up in $\langle x^2 \rangle$, and the parameter $\zeta$ falls out completely. Similarly, the second-order contact processes (where both the susceptible and infected travel) only show up in higher-order terms in the Taylor expansion, so we could have derived the same PDE using the simplified form of the contact distribution \eqref{eq:PR(IC)simplified}. 

The equations resemble a reaction-diffusion system with drift. This is made most explicit by re-expressing the differential equation in terms of $I$. 
\begin{equation}
    \frac{\partial I}{\partial t} =   \beta s I + s\div \left(D \grad I - \vec{v} I \right) - \gamma I \label{eq:iiC}
\end{equation}
Here, the diffusion coefficient and drift velocity are given by 
\begin{equation}
    D = \frac{\beta \langle x^2 \rangle}{d}
\end{equation}
and 
\begin{equation}
    \vec{v} = \frac{\beta \langle x^2 \rangle }{d}\grad \log N \label{eq:drift}
\end{equation}
respectively. The infection wave drifts toward higher populations, not because there are more people commuting to work in these places (that would be captured with $\zeta$), but simply because there are more people there capable of spreading infection. 

The diffusion coefficient for the susceptible and infected populations is the same since both arise from the same contact processes, so no Turing patterns can be observed.

Although the nonlinear PDE \eqref{eq:iiC} cannot be solved analytically, we can study the behavior when almost everybody is susceptible by setting $s=1$:
\begin{equation}
    \frac{\partial I}{\partial t} =   (\beta -\gamma) I + \div \left(D \grad I - \vec{v} I \right) . \label{eq:iiCLinear}
\end{equation}
We will focus on cases where there is initially some small number of people $I_0$ infected at the origin:
\begin{equation}
    I(\vec{x},0) = I_0 \delta^d (\vec{x}). \label{eq:InitialCondition}
\end{equation}
The infection density will immediately spread out and grow. There are two quantities which we will use to characterize this spreading process: the arrival time and the offset time. Although our analysis will be done in the continuum limit in terms of the infection density, these definitions can easily be modified for use with the discrete quantity $I_n$, so they can be calculated using real county-level data. The \textit{arrival time} $t_{\text{arrival}}$ at location $\vec{x}$ is the time it takes for $I(\vec{x},t)$ to exceed a critical value for the first time: $I(\vec{x},t_{\text{arrival}}) \geq I_c$. The \textit{offset time} $t_O$ is the smallest time which satisfies $I(0,t)\geq I_c$ and $\dot{I}(0,t)>0$. It is useful to think of $I_c$ as the infection density needed to detect the presence of the disease. Then, the arrival time $t_{\text{arrival}}(\vec{x})$ indicates how the disease is spreading at early times. The first location for the disease to be detected will typically be near the origin. The offset time is often equivalent to the arrival time at the origin except we also have to check that the infection density is increasing, since we could have $I(0,0)>I_c$, but $\dot{I}(0,0)<0$. Note that although the total number of infected individuals in the population $I_{\text{tot}}$ increases at early times whenever $R_0>1$, we will see that for our choice of initial condition, the infection density at the origin $I(0,t)$ will initially decrease for a time $\frac{d}{2}(\beta-\gamma)^{-1}$ before increasing. At early times, the infection spreads out in space before any one location experiences an outbreak. With our definition, the offset time gives us a metric for how long it will take for the first outbreak to occur. Understanding $t_O$ will reveal an interesting competition between the growth and diffusion processes in our model. 

\subsubsection{Solutions in an infinite domain}
\begin{figure}
    \centering
    \includegraphics[width=.45\textwidth]{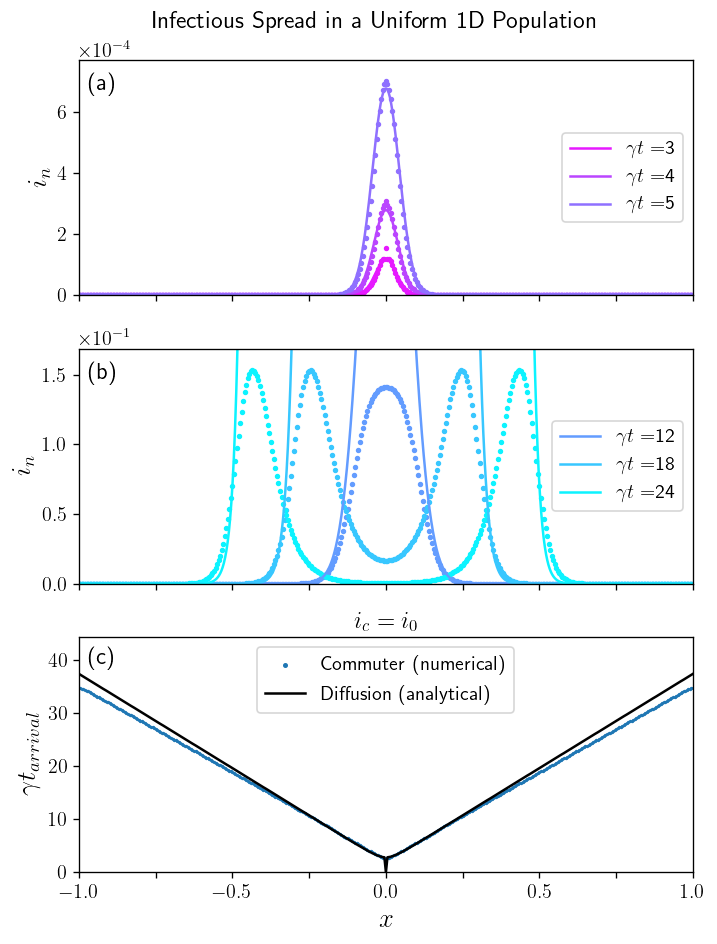}
    \caption{Infection waves traveling in $d=1$ from an initial infection at $x=0$ in a uniform population $N_n = $ constant. In (a) and (b), the fraction of people infected at each location $i_n$ is shown for different times represented by different colors. The points are numerical solutions to equation \eqref{eq:i} with $G(|x|)\sim e^{-|x|/l}$, and the solid lines are analytical solutions to \eqref{eq:I(x-vt,t)} multiplied by $dx$. Parameters used were $dx = .005$, $\sqrt{\langle x^2\rangle }=\sqrt{2l^2}=.01$, $i_0=10^{-4}$, and $R_0=2$. (a) At early times, $s\approx 1$, so our analytical result \eqref{eq:I(x-vt,t)} agrees with the numerical result. (b) For later times, the two solutions disagree, but \eqref{eq:I(x-vt,t)} still captures the propagation of the wave front. (c) The smallest time that solves $i_n(t_{\text{arrival}}) \geq i_c $ is displayed for each location.}
    \label{fig:1DInfectiousSpread}
\end{figure}

We begin our analysis by solving \eqref{eq:iiCLinear} on $\Omega=\mathbb{R}^d$ with boundary condition \eqref{eq:Neumann} and initial condition \eqref{eq:InitialCondition}. The solution for this case is just the well-known solution to the problem of diffusion with growth
\begin{equation}
    I(\vec{x},t) = I_0 \frac{\exp{-\frac{(\vec{x}-\vec{v}t)^2}{4Dt}}}{(4 \pi D t)^{d/2}} e^{(\beta - \gamma)t} . \label{eq:I(x-vt,t)}
\end{equation}
This exactly solves \eqref{eq:iiCLinear} when $\vec{v}(\vec{x})$ is constant, but for a slowly varying population, $\vec{v}(\vec{x})$ is also slowly varying, so we can approximate the solution by simply replacing $\vec{v}\rightarrow \vec{v}(\vec{x})$. Figure \ref{fig:1DInfectiousSpread}a shows that the numerical solution to \eqref{eq:i} agrees with \eqref{eq:I(x-vt,t)} at early times while \ref{fig:1DInfectiousSpread}b shows that the foot of the wave nearly matches the diffusion solution at all times. At long distances, there is some deviation from the diffusion solution due to the higher-order derivative terms we dropped. For all simulations in this section, an exponential commuting distribution was used. At early times, the infection diffuses from the source. At later times, Fisher waves with height $i_n \approx i_{\text{max}}$ (as given by \eqref{eq:imax}) can be observed traveling at constant speed; equality holds when $N$ is constant.

To understand why the linearized equations match the wave front of the nonlinear system, it is worth reviewing some results for pulled waves \cite{Ebert_Van_Saarloos_2000, Dieterle_Amir_2021}. Intuitively, the infection growth rate $(\beta s_n - \gamma) $ is maximized for $s_n=1$, so the dynamics are dominated by the behavior at the foot of the front where $i_n$ is small. Whenever the initial conditions decay in space faster than an exponential, the infection front will converge to a wave-like solution traveling at the Fisher wave velocity with corrections which decay in time \cite{Aronson_Weinberger_1978, Ebert_Van_Saarloos_2000}. For the simple initial condition \eqref{eq:InitialCondition}, we can demonstrate this convergence explicitly by calculating the arrival time at different locations.

To calculate the arrival time, we equate \eqref{eq:I(x-vt,t)} to $I_c$. There is no analytic form for $t_{\text{arrival}}(\vec{x})$, but it is easy enough to solve for $\vec{x}(t_{\text{arrival}})$:
\begin{align}
    \vec{x}(t_{\text{arrival}}) &= \vec{v}t_{\text{arrival}} \nonumber \\&\pm \vec{c}t_{\text{arrival}} \sqrt{1 -\frac{ \log( \left(4 \pi D t_{\text{arrival}}\right)^{d/2} \frac{I_c}{I_0})}{(\beta - \gamma)t_{\text{arrival}}}} \label{eq:x(t)}
\end{align}
where $\vec{c}=c\hat{r}$ and 
\begin{equation}
    c = \sqrt{4D(\beta-\gamma)}= \gamma \sqrt{\frac{4R_0(R_0-1)\langle x^2 \rangle}{d}} \label{eq:speed}
\end{equation}
is the Fisher wave speed. At small times, the expression in the square root may be imaginary, indicating that the disease has yet to be detected anywhere. At late times, the expression in the square root approaches 1, so in a uniform population, the infection front travels at a speed $\vec{v}+\vec{c}$ far from the source. Figure \ref{fig:1DInfectiousSpread}c shows the arrival time predicted by the commuter model and by the diffusion approximation. The diffusion approximation correctly predicts the offset time, and nearly predicts the speed. Corrections to the speed can be obtained by keeping higher order derivatives. The fact that the arrival time is linear in $x$, and the slope is independent of our choice of $i_c$ is a robust property of pulled waves. 

\begin{figure*}
    \centering
    \includegraphics[width=.8\textwidth]{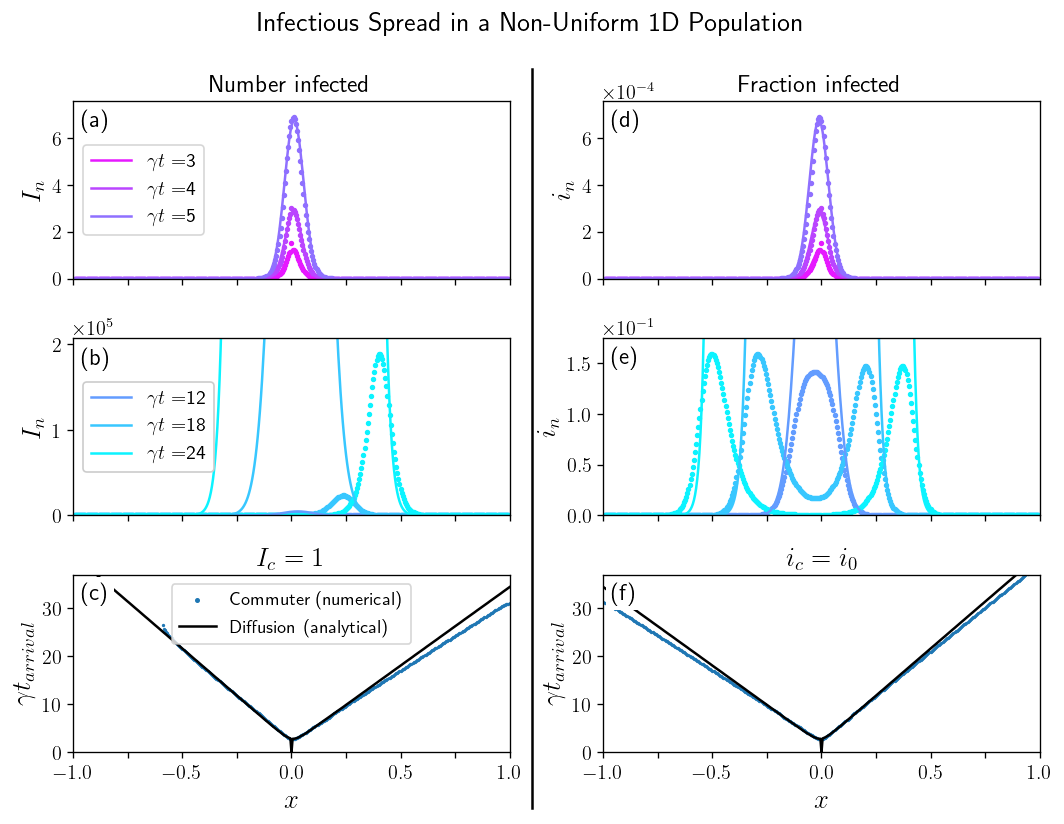}
    \caption{Infection waves traveling in $d=1$ from an initial infection at $x=0$ in a non-uniform population $N_n = N_0 e^{x_n/\lambda_N}$ such that the drift velocity \eqref{eq:drift} is constant. In (a) and (b), the number of people infected at each location $I_n$ is shown while in (c) and (d), the fraction of people infected at each location $i_n$ is shown. The points are numerical solutions to equation \eqref{eq:i} with $G(|x|)\sim e^{-|x|/l}$, and the solid lines are analytical solutions to \eqref{eq:I(x-vt,t)} multiplied by $dx$. Parameters used are $dx=.005$, $\sqrt{\langle x^2 \rangle}=\sqrt{2l^2}=.01$, $i_0=10^{-4}$, $R_0=2$, $N_0=10^4$, and $\lambda_N=.08$. (c) The smallest time that solves $I_n(t_{\text{arrival}}) \geq I_c $ is displayed for each location. Positive $x$ values have a smaller arrival time, indicating that $I_n$ drifts toward higher populations. (d) The smallest time that solves $i_n(t_{\text{arrival}}) \geq i_c $ is displayed for each location. Negative $x$ values have a smaller arrival time, indicating that $i_n$ drifts toward lower populations.}
    \label{fig:ExponentialPopulation}
\end{figure*}

When the population distribution is an exponential, the drift velocity is constant. This special case is shown in figure \ref{fig:ExponentialPopulation}. When plotted in terms of $I_n$, the infection drifts toward higher populations, reflecting the fact that there are more people capable of getting infected in more populated regions. When plotted in terms of $i_n$, the infection drifts toward lower populations, suggesting that a susceptible in a small town is initially at a higher risk of being infected than an individual in a city. However, even in this example with a highly non-uniform population, these drift effects are small, and the shape of the arrival time curves are similar regardless of whether we consider a critical infection count $I_c$ or a critical infection fraction $i_c$. For this simulation, several values of $\zeta$ were chosen with no noticeable differences in the results, consistent with the fact that the $\zeta$ corrections show up at fourth order in our Taylor expansion.

In addition to the drift velocity, there is an additional place at which $N$ can enter the calculations even in a constant population. Note that the arrival time in \eqref{eq:x(t)} satisfies $I(x,t_{\text{arrival}}) = I_c$, and we used the initial condition $I(x,0) = I_0 \delta(x)$. However, it may make sense for one or both of these conditions to be written in terms of $i$. If both the arrival condition and the initial condition are given in terms of $i$, then the factors of $N$ are eliminated. However, suppose one person is initially infected, but the infection is detected at a location when the fraction of cases exceeds a critical value $i(x,t)=i_c$. Then, the time offset will have an additional $\log N$ dependence, reflecting the fact that it takes longer for a disease to be detected in a larger population.

\begin{figure*}
    \centering    \includegraphics[width=.8\textwidth]{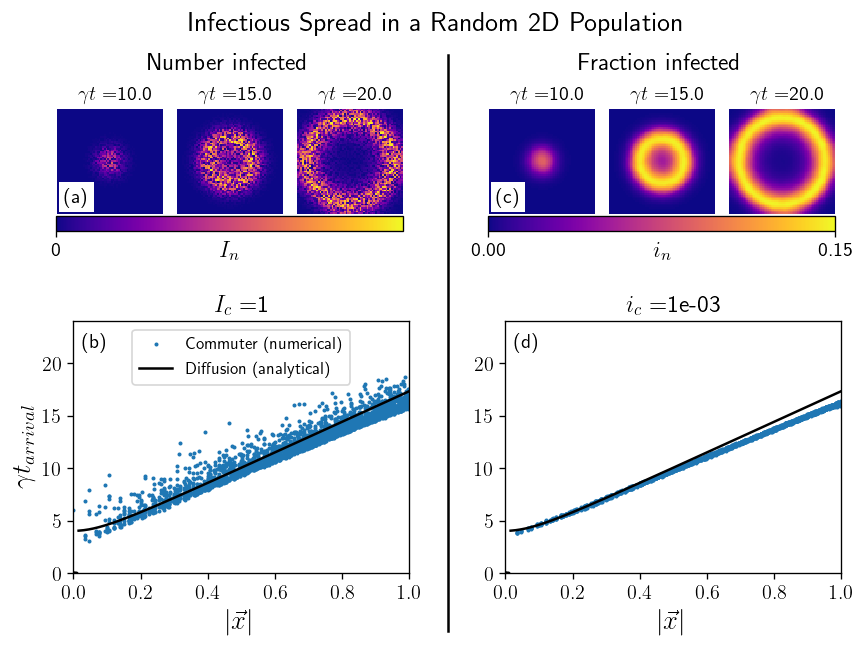}
    \caption{Infection waves traveling in $d=2$ from an initial infection at $\vec{x}=0$ in a non-uniform population with $N_n$ chosen from a uniform distribution between 0 and 2000. In (a), the number of people infected at each location $I_n$ is shown for three different times while in (c), the fraction of people infected at each location $i_n$ is shown for three different times. The points are numerical solutions to equation \eqref{eq:i} with $G(|\vec{x}|)\sim e^{-|\vec{x}|/l}$. The lattice is a 61 by 61 grid of side length $L=2$, so $dx=2/61$. Other parameters are $R_0=2$, $l=dx/2$, and $I_0=1$. (c) The smallest time that solves $I_n(t_{\text{arrival}}) \geq I_c $ is displayed for each location as a function of the distance from the origin $|\Vec{x}|$. (d) The smallest time that solves $i_n(t_{\text{arrival}}) \geq i_c $ is displayed for each location. In either case, the numerical solution is compared to the analytic solution assuming $N_n = 1000$. Notice that the noise observed when considering $I_n$ in (a)-(b) is eliminated when considering $i_n$ (c)-(d).}
    \label{fig:random2D}
\end{figure*}

So far, we have only considered long-wavelength population fluctuations. However, there is an opposite limit that is even simpler. When the population changes on the length scale of $\Delta x$, we empirically find that we can replace $N(\vec{x})$ with its average value and use the analytic results for the constant population case. See figure \ref{fig:random2D} for an example. This tells us that we can ignore the fact that the population of each county is different if we work in terms of $i(\vec{x},t)$ and focus only on large-scale population gradients.

Notice in these figures that the numerically calculated arrival time near the origin is greater than zero, indicating that there is some finite time before the infection is detected anywhere. At distances smaller than $\sim \sqrt{\langle x^2 \rangle}$, the analytical arrival time predicted by diffusion appears to be zero, a consequence of the highly localized initial conditions, but this is not a good indicator an outbreak is actually occurring close to the origin. It is clear from looking at these plots that there is a characteristic gap in the arrival time at the origin. We will quantify this gap by finding the offset time. For simplicity, we will focus on the constant population case where $\vec{v}=0$, so the first outbreak will occur at the origin. As a reminder, by outbreak, we mean two things: first, the infection density must exceed the critical value for detection $I(0,t) \geq I_c$, and second, the case counts must be increasing $\dot{I}(0,t)>0$. The offset time $t_O$ is the smallest time which satisfies both of these conditions. If $R_0\leq 1$, we can never satisfy the second condition because the infection density at the origin decreases monotonically, so $t_O$ does not exist. If $R_0>1$, the infection initially spreads out causing the infection density to decrease at the origin from its initially infinite value, but at later times exponential growth in the total number of cases causes the infection density at the origin to increase, so there exists a finite time $t^*$ at which the infection begins to increase. This places a lower bound on the offset time $t_O\geq t^*$. Solving $\dot{I}(0,t) =0$ reveals a particularly simple form for this time:
\begin{equation}
    t^*=\frac{d}{2}\frac{1}{\beta-\gamma}.
\end{equation}
Assuming $R_0>1$, $I(0,t)$ increases for times greater than $t^*$, but $I(0,t)$ may still be less than the critical density. In these cases, $I(0,t)$ falls from infinity to below $I_c$ at a time less than $t^*$, and grows above $I_c$ at a time greater than $t^*$. The time it takes to approach the critical density grows with the diffusion coefficient since this will lead to a broader spatial distribution of case counts at fixed total number of cases. It may be counter-intuitive that high mobility increases the arrival time, but recall that we are assuming the total number of contacts a person makes is fixed, and $D$ describes the spatial distribution of contacts. If contacts are only made between individuals in the same neighborhood (small $D$), a local outbreak will quickly occur, but if the same number of contacts are distributed across a state (large $D$), then it will take much longer for any particular location to detect a severe outbreak. In this way, the offset time should be thought of as the characteristic time it takes for the growth of the disease to dominate over the spread of the disease. 

Before continuing our analysis, it will be useful to introduce a dimensionless quantity $\bar{l}$ characterizing this competition between growth and spread
\begin{equation}
    \bar{l} \equiv \sqrt{\frac{d}{2e}}\sqrt{ \frac{4\pi D}{\beta-\gamma}} \left(\frac{I_c}{I_0}\right)^{1/d} =  \sqrt{\frac{2\pi}{e}}\sqrt{ \frac{R_0 \langle x^2 \rangle }{R_0-1}} \left(\frac{I_c}{I_0}\right)^{1/d} .
\end{equation}
Rewriting equation \eqref{eq:I(x-vt,t)} in terms of $\bar{l}$,
\begin{equation}
    \left(\frac{I(0,t)}{I_c}\right)^{2/d} =  \frac{\exp{t/t^*}}{e\bar{l}^2 t/t^*}.\label{eq:I(0,t)}
\end{equation}
When we have $t_O > t^*$, the two solutions to $I(0,t) = I_c$ are given by solving the following transcendental equation:
\begin{equation}
 -\frac{t}{t^*} e^{-t/t^*} = - \frac{1}{e\bar{l}^2} \label{eq:tOffInfinite}.
\end{equation}
When $0<\bar{l}<1$, there is no real solution to the above equation, and we are in the regime where diffusion is small enough that $I(0,t)$ never falls below $I_c$. When $\bar{l}>1$, there are two solutions (which converge to one solution for the special case $\bar{l}=1$):
\begin{equation}
    t = \left\{-t^* W_{0}\left(-\frac{1}{e\bar{l}^2}\right), -t^* W_{-1}\left(-\frac{1}{e\bar{l}^2}\right) \right\}. \label{eq:Lambert}
\end{equation}
Here, $W_k$ denotes the Lambert W-function and $k$ labels the branch cut. The first solution is the time at which $I(0,t)$ falls below $I_c$. The second solution is the offset time. Now, we can explicitly write down the offset time for all values of $\bar{l}$:
\begin{equation}
    t_O = \begin{cases} t^* , &\bar{l} \in (0,1) \\ 
     -t^* W_{-1} \left(-\frac{1}{e\bar{l}^2} \right), &  \bar{l} \in [1,\infty) .
    \end{cases} \label{eq:tOffsetExplicit}
\end{equation}
\begin{figure*}
    \centering
    \includegraphics[width=.8\textwidth]{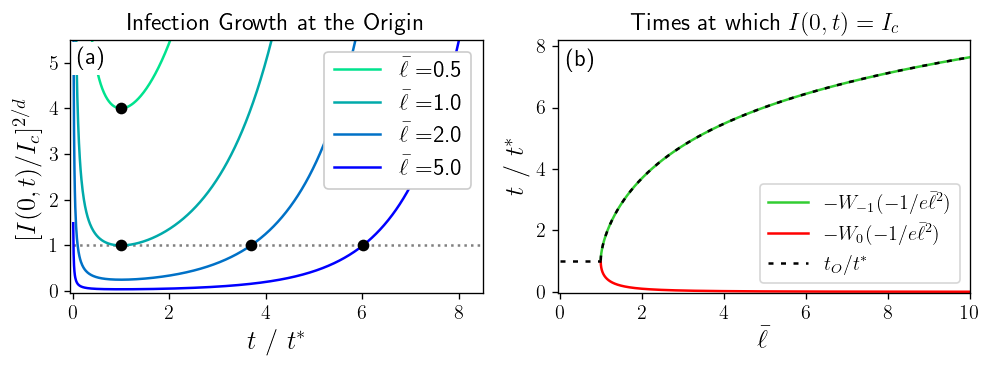}
    \caption{(a) Solutions to \eqref{eq:I(0,t)} for different choices of $\bar{l}$ (colored lines). The dashed gray line is the line $I(0,t) = I_c$, and the black dots indicate the offset time. (b) The red and green curves are the two solutions to equation \eqref{eq:Lambert}, while the black dashed line is the normalized offset time \eqref{eq:tOffsetExplicit}. For $\bar{l}<1$, $t_O=t^*$ which is the smallest time at which $\dot{I}(0,t)>0$. For $\bar{l}>1$, $t_O=-t^* W_{-1}(-1/e\bar{l}^2)$. }
    \label{fig:Lambert}
\end{figure*}
Looking at the form of $\bar{l}$, one can conclude that a reasonable choice of $\bar{l}$ will be greater than one. To see this, note that, $\bar{l}^d >  \sqrt{\langle x^2 \rangle}^d I_c/I_0$ which for $I_0=1$ can be interpreted as there being $I_c$ sick individuals within the characteristic interaction volume. If this number were less than one, then the disease would not be detected, and this would be an inappropriately small choice of $I_c$. Therefore, for any reasonable definition of $I_c$, we will be in the regime $\bar{l}>1$. When $\bar{l}>1$, the offset time is inversely proportional to $(R_0-1)$ and grows monotonically with $\bar{l}$ as seen in figure \ref{fig:Lambert}.

Now that we know there is a finite time offset greater than $t^*$ for reasonable choices of $\bar{l}$, we can expand the arrival time about the offset time, and study the observed spatial spreading patterns close to the origin. Plugging $t_{\text{arrival}} = t_O+\delta t$ into \eqref{eq:x(t)} and expanding to linear order in $\delta t$ gives
\begin{equation}
    \delta t = \frac{1}{t_O-t^*} \left(\frac{|\vec{x}|}{c} \right)^2 .
\end{equation}
We also know that at large distances, the arrival time is $|\vec{x}|/c$, and $c\sim \sqrt{\langle x^2 \rangle}$. Using these two limiting behaviors, we can estimate the arrival time at all distances from the origin:
\begin{equation}
    t_{\text{arrival}}(\vec{x}) =t_O - \frac{t_O-t^*}{2}+\sqrt{\left( \frac{t_O-t^*}{2}\right)^2 + \left(\frac{|\vec{x}|}{c} \right)^2} . \label{eq:tarrival}
\end{equation}
Close to the origin, the infection is observed to spread diffusively, but far from the origin, waves form. The characteristic distance over which the spread transitions from diffusive to ballistic is $c(t_O-t^*)/2$. Interestingly, the arrival time close to the origin (as characterized by $t_O$) increases with $\sqrt{\langle x^2 \rangle}$, but the arrival time far from the origin decreases with $ \sqrt{\langle x^2 \rangle} $. Thus, long-distance commutes delay an outbreak in any one location because the interaction radius is increased, keeping the infection density everywhere low. However, once the exponential growth takes over, these low infection-density regions each spawn a local outbreak, and the disease is quickly detected throughout the whole domain. In the opposite limit, when commutes are short range, social mixing leads to a high infection density causing a quick outbreak at the origin, but the propagation speed of the infection is slow, so it takes a while for the disease to be detected far from the origin. 

\subsubsection{Solutions in a finite domain}
When our domain is bounded $\Omega = [-\frac{L}{2},\frac{L}{2}]^d$, we will neither be able to assert that $\langle x^\mu \rangle = 0$ nor that $\langle x^2 \rangle$ is independent of position, but we can attempt to understand the boundary effects by using equation \eqref{eq:iiCLinear} along with Neumann boundary conditions \eqref{eq:Neumann}. Physically, these boundaries could correspond to bodies of water or national borders. 

If $L$ is large enough and \eqref{eq:Mollison} is satisfied, we expect to see infection waves, and the methods of the previous section can be used, but, if $L$ is small, the infection quickly reaches the boundary before a wave can form, and the homogeneous dynamics \eqref{eq:SIR} are recovered. We will only consider a uniform population in this section.  Solving the linearization of \eqref{eq:iiC} with boundary condition \eqref{eq:Neumann} and initial condition \eqref{eq:InitialCondition} gives
\begin{align}
    I(\vec{x},t) = \frac{I_0}{L^d}  \prod_{\mu=1}^d  \sum_{n_{\mu}=-\infty}^\infty &\cos(\frac{2\pi n_\mu x^\mu}{L}) \nonumber\\ \times & \exp \left(-D \left(\frac{2\pi n_\mu}{L}\right)^2 t \right) e^{(\beta-\gamma)t}.
\end{align}
In particular, for $\vec{x}=0$ we get
\begin{equation}
    I(0,t) = \frac{I_0}{L^d}   \left(\sum_{n=-\infty}^\infty  e^{-D (\frac{2\pi n}{L})^2 t} \right)^d e^{(\beta-\gamma)t}.
\end{equation}
The spatial dependence decays away over a time $L^2 /4\pi^2 D$, so if the arrival time is everywhere larger than or comparable to this value, the spread will appear to be homogeneous. It's worth mentioning that for $L^2 /4\pi^2 D$ large, the terms in the sum change slowly, so one can turn the sum into an integral and recover equation \eqref{eq:I(x-vt,t)}. 

In the previous section, we introduced the dimensionless commuting length $\bar{l}$ which parameterized the offset time in an infinite domain. Here we would like to introduce a second dimensionless length scale $\bar{L}$ to characterize the system size:
\begin{equation}
     \bar{L} \equiv L \left(\frac{I_c}{I_0} \right)^{1/d}.
\end{equation}
As mentioned in the previous section, reasonable choices of $I_c$ will lead to the $\dot{I}(0,t)>0$ condition automatically being satisfied (and a finite domain only reduces the time it takes to satisfy this condition), so let's assume we are at times after which $\dot{I}(0,t)>0$.  Then the offset time is given by solving $I(0,t_O)=I_c$:
\begin{equation}
    \bar{L}^2 =  \left(\sum_{n=-\infty}^\infty  e^{-\pi e n^2 (\bar{l}^2 / \bar{L}^2)  t_O/t^*}\right)^{2} e^{t_O/t^*}. \label{eq:tOGeneral}
\end{equation}
Equation \eqref{eq:tOffInfinite} is recovered in the small $\Bar{l}/\bar{L}$ limit where the sum can be converted to an integral. In the opposite limit of a well-mixed population (where we only keep the $n=0$ term), the offset time is simply
\begin{align}
     t_O^{WM}\equiv t_O(\bar{l}\gg \bar{L}) = \frac{1}{\beta - \gamma } \log (\frac{I_c L^d}{I_0} ).\label{eq:tOffsetWM}
\end{align}
That is, a larger system (at fixed critical infection density) will have a larger offset time since there is more space for the infection to spread out. This limit gives the same result as the well-mixed population \eqref{eq:tArrTot}. In a well-mixed population, the entire population reaches $I_c$ simultaneously, at which point $I_c L^d$ people are infected, which is necessarily greater than the initial number of people infected $I_0$, so we must have that $\bar{L}>1$, and $t_O^{WM}>0$.

\begin{figure*}
    \centering
    \includegraphics[width=.45\textwidth]{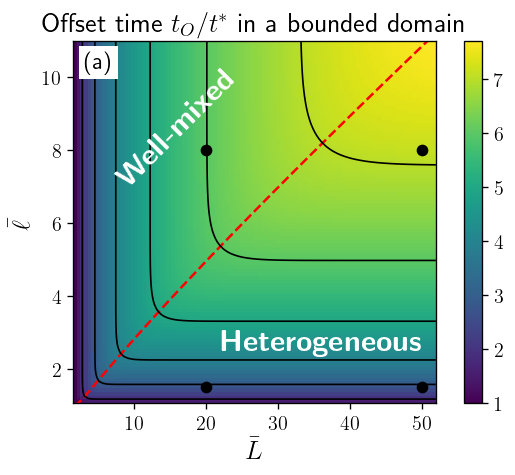}
    \includegraphics[width=.5\textwidth]{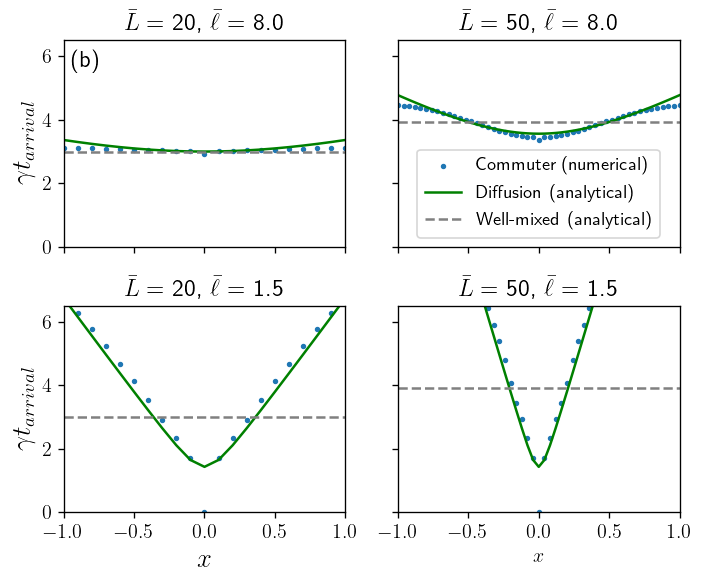}
    \includegraphics[width=.9\textwidth]{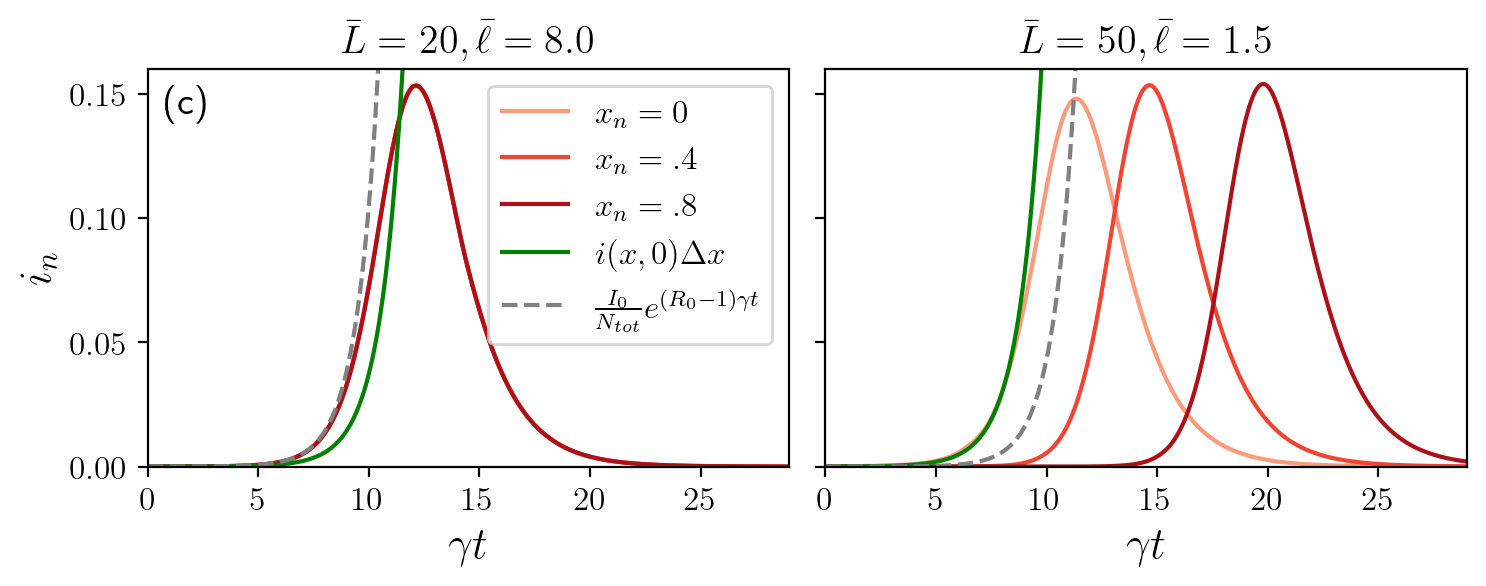}
    \caption{Demonstration of finite size effects on arrival time dynamics. (a) The theoretical time offset \eqref{eq:tOGeneral} is shown as a function of $\bar{L}$ and $\bar{l}$. In the top-left region, the contours are nearly vertical, indicating that the offset time only depends on system size, and \eqref{eq:tOffsetWM} can be used; in the bottom-right region, the contours are nearly horizontal, indicating the offset time is independent of system size, and \eqref{eq:tOffsetExplicit} can be used. The dashed red line corresponds to equation \eqref{eq:lStar}, where the well-mixed and $\bar{L}\rightarrow \infty$ solutions agree. In (b), the arrival time was found numerically by solving \eqref{eq:i} with $G(|\vec{x}|)\sim e^{-|\vec{x}|/l}$ in a uniform 1D population of size $L=2$, for different choices of $\bar{L}$ and $\bar{l}$ corresponding to the black dots in (a). $\bar{L}$ was varied by changing the step size while keeping $I_c/I_0$ fixed to $1/dx$. $\bar{l}$ was varied by changing the commuting length $l$. The green lines correspond to the analytical result in an infinitely large population satisfying \eqref{eq:tarrival} while the gray dashed lines give the well-mixed result \eqref{eq:tOffsetWM}. (c) Infection curves at different locations for an epidemic in the well-mixed regime (left) and an epidemic in the heterogeneous regime (right). The green line shows the diffusion result \eqref{eq:I(x-vt,t)}, and the gray dashed line shows the solution to the linearization of \eqref{eq:sIr}.}
    \label{fig:FiniteSizeEffects}
\end{figure*}

One can check numerically that $t_O/t^*$ is an increasing function of both $\bar{l}$ and $\bar{L}$. For moderate values of $\bar{L}$, the offset time resembles \eqref{eq:tOffsetExplicit} for small values of $\bar{l}$ and resembles \eqref{eq:tOffsetWM} for larger values of $\bar{l}$. Figure \ref{fig:FiniteSizeEffects}a demonstrates that the offset time can typically be estimated by one of these two simplified expressions, giving us a simple way of characterizing the spreading dynamics as either heterogeneous or homogeneous (well-mixed). By equating these two asymptotic expressions, we find that the transition occurs when
\begin{equation}
    \log (\bar{L}^2) = -W_{-1}\left(-\frac{1}{e\bar{l}^2}\right) \implies \bar{l}^* = \sqrt{\frac{1}{e}\frac{\bar{L}^2}{\log (\bar{L}^2)}} \label{eq:lStar}
\end{equation}
That is, for $\bar{l}<\bar{l}^*$, the disease dynamics close to the origin can be approximated by those of the infinite system, and for $\bar{l}>\bar{l}^*$, the arrival time everywhere is approximated by the well-mixed result. Some sample arrival times calculated by solving \eqref{eq:CommuterModelsir} are shown in figure \ref{fig:FiniteSizeEffects}b and compared to the diffusion \eqref{eq:tarrival} and well-mixed results \eqref{eq:tOffsetWM}. Note that we could have important boundary effects at positions away from the origin even when  $\bar{l}<\bar{l}^*$. Some sample infection curves are shown in figure \ref{fig:FiniteSizeEffects}c. In the well-mixed regime, the infection curves at different locations follow \eqref{eq:linearizedSIR} with $i_{\text{tot}}(0) = I_0/N_{\text{tot}}$. In the heterogeneous regime, the infection curves separate in space, and the infection curve at the origin follows the diffusion model \eqref{eq:I(x-vt,t)}. One proposed way to capture the heterogeneity in disease spread is via an entropy function which measures the similarity of these  infection curves at different locations \cite{Colizza_Barrat_Barthelemy_Vespignani_2006}. However, this quantity requires knowledge of $i_n(t)$  for all $n$ and $t$. What we have demonstrated is that simply looking at one location in space (the origin) at early times also characterizes the spreading behavior.

Even though we have given much attention to the offset time in cases where the diffusion approximation holds, we now understand the two most important limits. The smallest offset time corresponds to the case of a population with vanishingly small commuting range such that the infection density can quickly reach the threshold value. The largest offset time corresponds to a well-mixed population. Although $\bar{L}$ is defined for all commuting distributions, $\bar{l}$ is only defined for localized commuting distributions. What, then, do we expect for delocalized commuting distributions? The infection quickly reaches the boundary due to the divergent moments, so the more spread out the commuting distribution is, the closer the offset time will be to the well-mixed prediction. In general, the well-mixed result gives an upper bound to the offset time: $t_O \leq t_O^{WM}$.

\subsection{Localized commuting beyond the diffusion approximation}
\subsubsection{Exact calculation of the wave speed}
When truncating our Taylor series to quadratic order, we found a reaction-diffusion system. This should work well at small values of $x$, but at larger values of $x$, the higher-order derivatives will, in general, matter. Therefore, we might expect that the calculation of the time offset (which is defined at the origin) should work well under the diffusion approximation provided $\langle x^2 \rangle$ is finite. However, the wave speed should be calculated far from the origin, where the diffusion approximation is no longer expected to work. Although the diffusion approximation correctly predicts the existence of waves, we need a new tool to quantitatively predict the wave speed for a given contact distribution. Mollison showed that for the special case of an exponential contact distribution, the exact wave speed can be calculated \cite{Mollison_1972b} and has a noticeable deviation from the result obtained from the diffusion approximation. Later, Daniels generalized the result to show how to calculate the wave speed for any localized contact distribution by using moment-generating functions \cite{Daniels_1975}. Here, we will apply the technique of moment-generating functions to the commuter model. We have a few probability distributions to choose from, but the most relevant for the dynamics will be the commuter distribution $G(|\Vec{x}|)$ and the contact distribution $\beta(\vec{x},\vec{y})$. Consider two related quantities:
\begin{equation}
    \phi(\vec{k}) = \int e^{\vec{k}\cdot \vec{x}} G(|\vec{x}|) d^dx 
\end{equation}
\begin{equation}
    \psi(\vec{k},\vec{x}) = \beta^{-1}\int e^{\vec{k}\cdot (\vec{y}-\vec{x})} \beta(\vec{x},\vec{y}) d^dy.
\end{equation}
When the population is uniform, $\psi(\vec{k},\vec{x})=\psi(\vec{k})$, and we have the simple relationship $\psi(\vec{k}) = \phi(\vec{k})^2$. 
Plugging the ansatz $i(\vec{x},t)\sim e^{\vec{k}\cdot (\vec{x}-\vec{c}t)}$ into \eqref{eq:iC} with $s(\vec{x},t)=1$ and assuming a uniform population gives a constraint on $c(\vec{k})$:
\begin{equation}
    c(\vec{k}) = \frac{\beta\phi(\vec{k})^2-\gamma}{|\vec{k}\cdot \hat{r}|}. \label{eq:speedCorrected}
\end{equation}
For a localized initial condition like we have been considering, the correct choice of $\vec{k}$ is that which minimizes $c(\vec{k})$. Note that the diffusion approximation considered previously amounts to keeping only quadratic terms in our moment-generating function. If we make the approximation $\phi(k) \approx 1+\frac{1}{2d}k^2 \langle x^2 \rangle \implies \phi(k)^2 \approx 1+\frac{1}{d}k^2 \langle x^2 \rangle $, then one can easily check that the minimum wave speed is that given in equation \eqref{eq:speed}. In order to find the non-perturbative wave speed, one must first find the exact moment-generating function $\phi(\Vec{k})$ and then find the minimum of \eqref{eq:speedCorrected} numerically.

\begin{figure*}
    \centering
    \includegraphics[width=.965\textwidth]{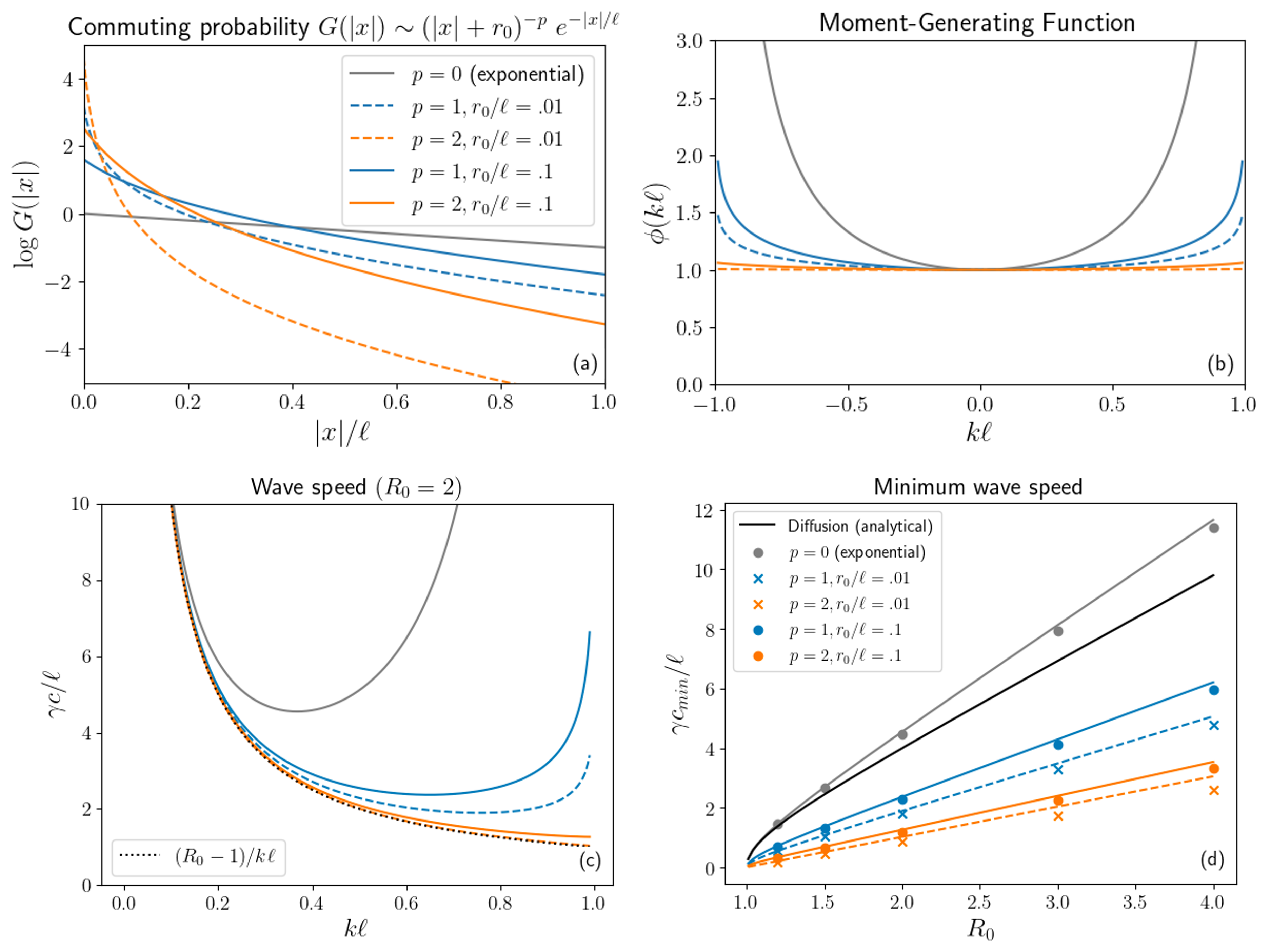}
    \caption{Demonstration of how to calculate the infection wave speed of a general localized contact distribution. (a) Truncated power law commuting distributions and (b) their corresponding moment-generating functions as given by equation \eqref{eq:phiTPL} for various powers $p$ and length scale ratios $r_0/l$. (c) Wave speed for each $kl$ at a fixed $R_0$, as predicted by equation \eqref{eq:speedCorrected}. The physical wave speed is the minimum speed on this curve. (d) Curves are generated by minimizing \eqref{eq:speedCorrected} over all $kl$ for different choices of $R_0$. The circles and crosses denote values obtained numerically by finding the slope of the arrival time far from the origin. The numerical results are compared to the diffusion result \eqref{eq:speed} shown with a black line . }
    \label{fig:MGF}
\end{figure*}

As an example, let's find $\phi(\vec{k})$ and $c(\vec{k})$ in 1D when $G$ takes the form of equation \eqref{eq:TruncatedPowerLaw}. For $|kl|<1$,
\begin{align}
    \phi(kl) &= (1+kl)^{p-1}e^{+klr_0/l }\frac{\Gamma\left(-(p-1),(1+kl)r_0/l \right) }{2\Gamma \left(-(p-1),r_0/l\right)}\nonumber\\ &+(1-kl)^{p-1}e^{-klr_0/l } \frac{\Gamma\left(-(p-1),(1-kl)r_0/l \right)}{2\Gamma \left(-(p-1),r_0/l\right)}\label{eq:phiTPL}
\end{align}
where $\Gamma(s,x)$ denotes the incomplete Gamma function 
\begin{equation}
    \Gamma(s,x) = \int_x^{\infty} w^{s-1} e^{-w}dw.
\end{equation}
Taking the limit $r_0/l \rightarrow 0$ is difficult, especially when $p$ is a positive integer, but intuitively, as $r_0/l \rightarrow 0$, $G(|\vec{x}|)$ becomes sharply peaked at $\vec{x}=0$, so $\phi(kl)\approx 1$. Although the $kl$-dependence is important if we want to find any nonzero moments, taking $\phi(kl)=1$ actually gives us a good estimate for $c$. In that case, the minimum value of $c$ occurs when $|kl|=1$ (a strong indication that our diffusion approximation will fail), and the minimum speed is equal to $(R_0-1)l$, unlike that predicted by equation \eqref{eq:speed}. Unsurprisingly, the speed becomes infinite for a power law ($l=\infty$) since we no longer satisfy equation \eqref{eq:Mollison}. Setting $\phi(kl)=1$ fails when $p\ll 1$ where the distribution becomes exponential and $\phi(kl) \approx \frac{1}{2}[(1+kl)^{p-1} +(1-kl)^{p-1}]$. Notably, when $p=0$ (exponential distribution), we have $\phi(kl)=\frac{1}{2}[1-(kl)^2]^{-1}$, and $c$ is slightly greater than that predicted by the diffusion approximation, consistent with the small discrepancy in the arrival times shown in figure \ref{fig:1DInfectiousSpread}. The moment-generating functions and wave speeds for various choices of $p$ and $r_0/l$ are shown in figure \ref{fig:MGF}. When a truncated power law with a small length scale is used, there is an even larger disagreement with the diffusion approximation. The key takeaway is that even though the Taylor series is valid for any localized commuting distribution, keeping only the quadratic terms fails to predict the correct wave speed, and higher moments should be considered. Even more effects absent from the reaction-diffusion system can be found when $N(\Vec{x})$ is nonuniform, as will be explored in the next section. 

\begin{figure*}
    \centering
    \includegraphics[width=.9\textwidth]{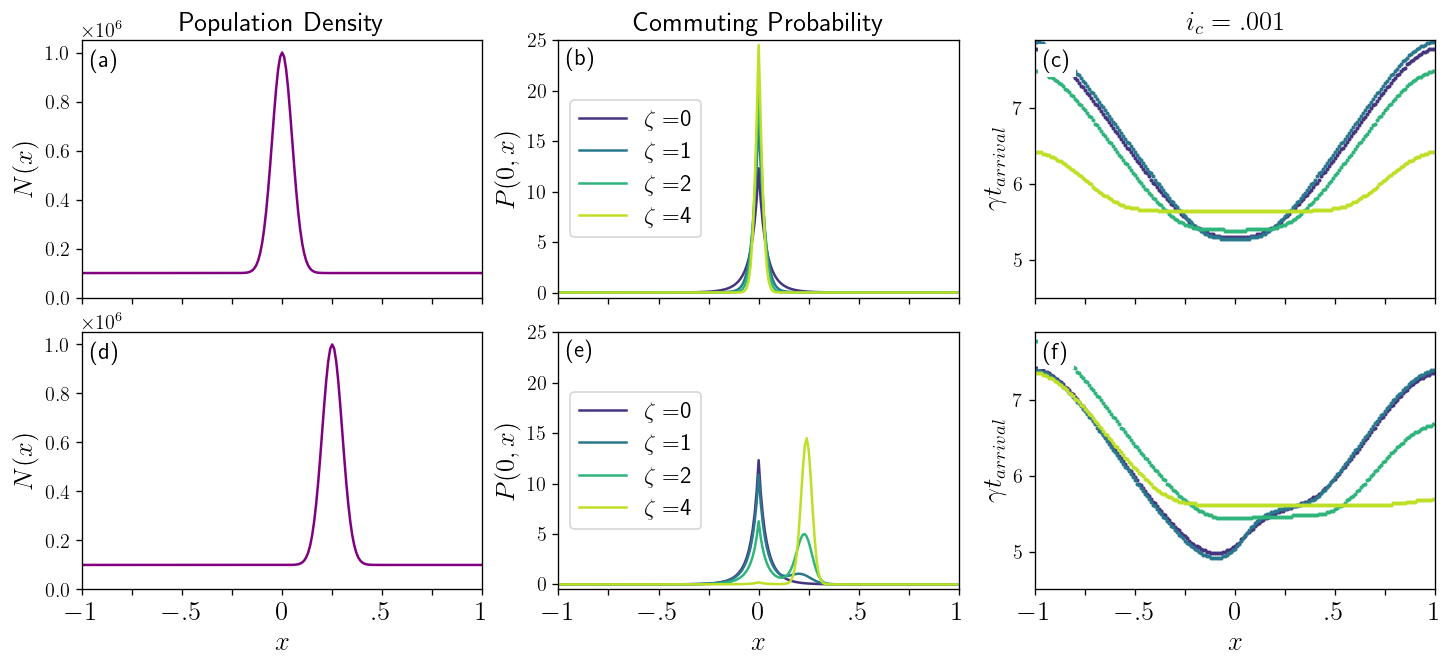}
    \caption{Demonstration of the $\zeta$-dependence in the continuous commuter model. (a) A Gaussian population profile was chosen whose center lies at the origin: $N(x) = 10^5+9\cross 10^5 e^{-x^2/2\lambda_N^2}$. (b) Commuting probabilities of the form $P(0,x) \sim N(x)^\zeta e^{-|x|/l}/(|x|+r_0)^2$. The probability for an individual living at the origin to travel to the city is larger for higher $\zeta$, reducing the arrival time near the city. (c) Arrival times obtained by numerically solving \eqref{eq:i}. The presence of the city leads to more homogeneous dynamics when $\zeta$ is increased. Parameters used were $l=.1$, $r0=.1, \lambda_N=.05, R_0=2$, $I_0=1$, $dx=.01$, $L=2$. In (d)-(f), the simulation is repeated except the city is shifted to the right: $N(x) = 10^5+9\cross 10^5 e^{-(x-.25)^2/2\lambda_N^2}$. In either case, the infection was initialized at the origin.}
    \label{fig:fourthOrder}
\end{figure*}

\subsubsection{Fourth-order corrections to the reaction-diffusion system}
Returning to the calculation performed to obtain the reaction-diffusion system, here we will instead Taylor expand $i(\vec{z},t)$ and $N(\vec{z})$ to fourth order. The calculation is tedious, so we only give the result for 1D, where primes denote spatial derivatives, and all quantities on the right-hand side are evaluated at $x$.
\begin{widetext}
\begin{align}
    \int d y & \hspace{.1cm}P(x,y) \frac{\int d z \hspace{.1cm}P(z,y) N(z) i(z,t) }{\int d^d z \hspace{.1cm}P(z,y)  N(z)}
    \approx i + \Bigg[i''  +\frac{N'}{N} i' \Bigg]\langle x^2 \rangle \nonumber \\
    &+\Bigg[\frac{1}{12} i''''+\frac{1}{6}\frac{N'}{N}i''' + \left(\frac{1}{2}\zeta(\zeta-1)\frac{N'^2}{N^2}+\frac{1}{4}\frac{N''}{N}\right)i'' \nonumber \\
    &\hspace{.4cm}+ \left(-\frac{1}{2}\zeta(\zeta-1)\frac{N'^3
    }{N^3}+\zeta(\zeta-1)\frac{N' N''}{N^2}+\frac{1}{6}\frac{N'''}{N} \right)i'\Bigg]\langle x^4\rangle \nonumber\\
    &+\Bigg[\frac{1}{4}i'''' +\frac{1}{2}\frac{N'}{N}i''' + \left((-\frac{5}{4}\zeta^2 + \frac{9}{4}\zeta - 1)\frac{N'^2}{N^2}+(1-\zeta)\frac{N''}{N}\right)i'' \nonumber\\ &\hspace{.4cm}+ \left((1-\zeta)(1-2\zeta)\frac{N'^3}{N^3}+(-4\zeta^2+\frac{13}{2}\zeta-2)\frac{N'N''}{N^2}+(-\zeta+\frac{1}{2})\frac{N'''}{N}\right)i'\Bigg] \langle x^2 \rangle^2 \label{eq:fourthorder}
\end{align}
\end{widetext}
The first line is the diffusion result, but the numerous other terms are new. The higher-order $N$-independent terms correct the wave speed as shown in the last section. The $\langle x^2 \rangle^2$ terms involve second-order processes which we would not have obtained by considering only dynamics where either the susceptible or the infected travel. We now see terms which depend on $\zeta$. While this equation is too complicated to say exactly how $\zeta$ changes quantities like the arrival time, we can see a few things. Because the $\zeta$ terms all involve at least two derivatives of $N$ and factors of the characteristic commuting distance to the fourth power, in order for these terms to be non-negligible, we need $N$ to change on length scales comparable to the characteristic commuting length. If the population changes on much greater length scales, then the dominant effect will be the $N'/N$ term (which may also be negligible). If the population changes on much smaller length scales, then the population fluctuations can be averaged over as already shown. See figure \ref{fig:fourthOrder} for an example population where the $\zeta$-dependence is important. First, notice that larger values of $\zeta$ lead to more homogeneous dynamics, a phenomenon which will be explained by equation \eqref{eq:zetaInfinity}. Next, notice in the bottom row that although the drift term would cause the infection to travel left toward lower populations (when plotted in terms of $i_n$), the higher-order terms drive the infection toward the city for larger values of $\zeta$. Finally, notice that the arrival times for $\zeta = 0$ and $\zeta =1$ are nearly indistinguishable. Looking at the above equation, we see that all of the terms proportional to $\langle x^4 \rangle$ are identical for $\zeta=0$ and $\zeta=1$. It is not clear why we have this symmetry, but this manifests itself in our simulations where it is often hard to distinguish between commuting populations with $\zeta=0$ and $\zeta=1$.

\begin{figure*}
    \centering
    \includegraphics[width=.75\textwidth]{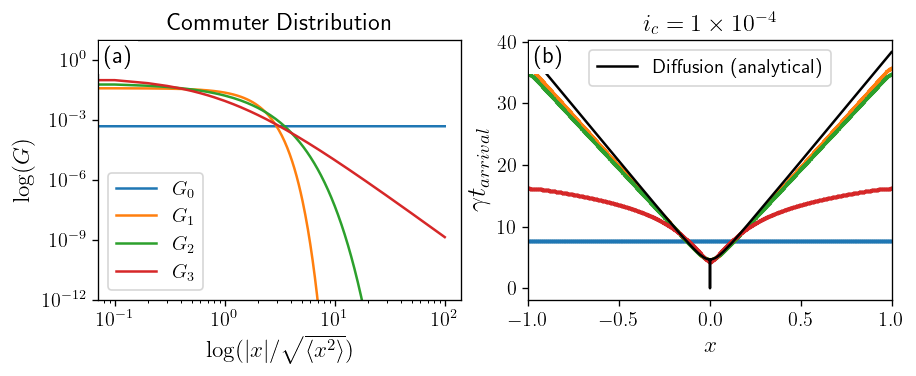}
    \includegraphics[width=.75\textwidth]{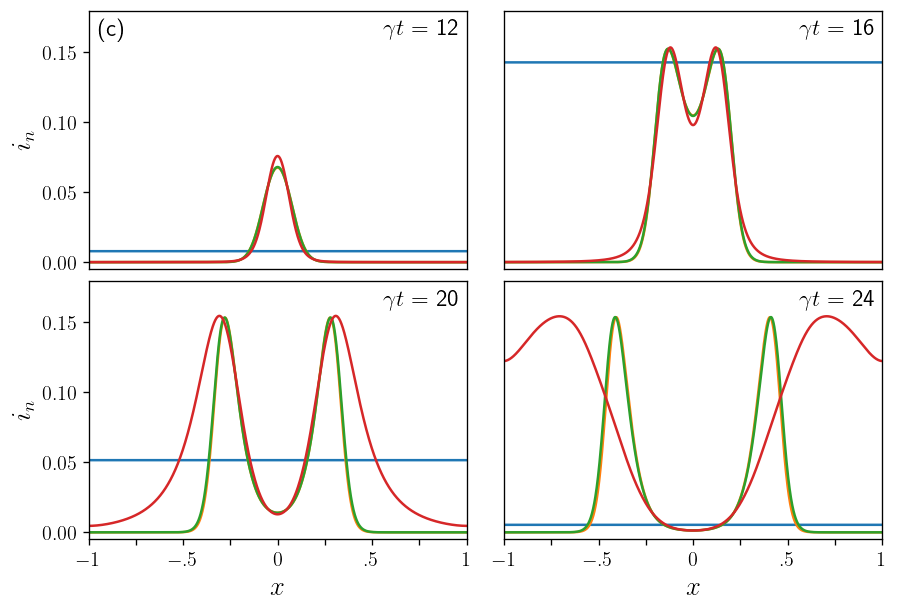}
    \caption{Numerical results for the 1D homogeneous commuter model for four different choices of $G$: completely delocalized $G_0(x)=\frac{1}{L}$, Gaussian $G_1(x)=\frac{e^{-x^2/2\langle x^2 \rangle}}{\sqrt{2\pi \langle x^2 \rangle}}$, exponential $G_2(x) = \frac{e^{-|x|/\sqrt{\langle x^2 \rangle /2}}}{2\sqrt{\langle x^2 \rangle/2 }}$, and power law $G_3(x) = \frac{3}{2\sqrt{\langle x^2 \rangle} }\left(1+\frac{|x|}{\sqrt{\langle x^2 \rangle}}\right)^{-4}$. (a) Commuter distributions plotted on a log scale. Note that the value of $\sqrt{\langle x^2\rangle}$ was fixed to be the same for $G_1$, $G_2$, and $G_3$. (b) The arrival times for each commuter distribution were found. The Gaussian and exponential distributions give rise to a linear arrival time far from the origin, consistent with the diffusion result shown in black. The power law initially resembles the diffusion result, but becomes sublinear far from the origin. Despite having the largest time offset, the completely delocalized distribution has the smallest arrival time far from the origin. (c) Four time snapshots of the infection dynamics for each commuter distribution. The power law displays an accelerating infection front. Parameters used were $dx=.001, \sqrt{\langle x^2 \rangle} =.01$ (for $G_1, G_2, G_3$), $L=2$, $i_0=10^{-4}$, and $R_0=2$. }
    \label{fig:GComparison}
\end{figure*}

\begin{figure*}
    \centering
    \includegraphics[width=.8\textwidth]{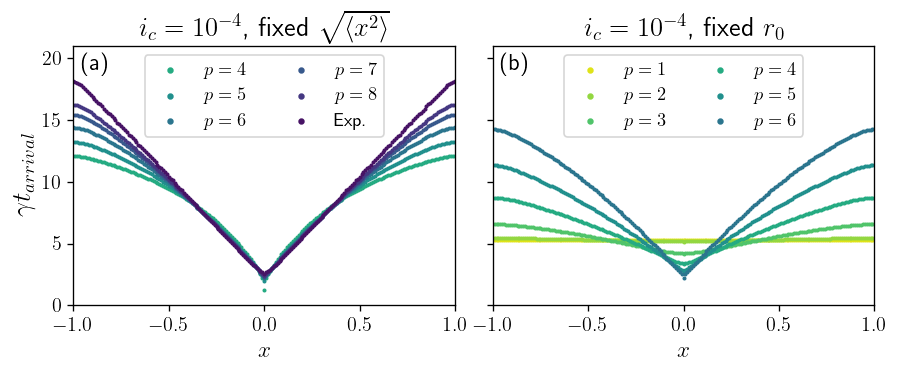}
    \caption{Arrival times for different power law commuter distributions $G(x)\sim (|x|+r_0)^{-p}$. In (a), $r_0$ was chosen for each $p$ to fix $\sqrt{\langle x^2 \rangle} = .02 $ using $r_0(p)=\sqrt{(p-2)(p-3)\langle x^2 \rangle /2}$. In (b), $r_0=.05$ was fixed. Additional parameters used were $dx=.01$, $i_0=.0001$, and $R_0=2$.}
    \label{fig:ArrivalTime(r0)}
\end{figure*}

\subsection{Continuous commuter model results: Most general case}
The previous two sections have focused on two extreme limits: if the system is small or commuters travel long distances, then the well-mixed dynamics \eqref{eq:SIR} are recovered, but if the travel probabilities fall off exponentially and the system size is large enough, a reaction-diffusion system approximates the dynamics. In either of these two limits, the precise form of the commuting probabilities is unimportant (only $\langle x^2\rangle$ matters to leading order). Our first indication that the precise commuting patterns matters shows up at fourth order in our Taylor series, as shown in the previous section. This hints at the fact that the most interesting dynamics happen for commuting distributions which are delocalized so the reaction-diffusion system fails but also not too long-ranged as to lead to homogeneous dynamics. So far, we have only studied cases where \eqref{eq:Mollison} is satisfied such that our Taylor series is valid (even if it may require us to go to higher-orders). In this section, we relax that assumption and study the role of $\zeta$ and $G$ for general commuting probabilities $P(\vec{x},\vec{y})\sim N(\vec{y})^\zeta G(|\vec{x}-\vec{y}|)$.

Let's first study the role of $G$ in a uniform population. See figure \ref{fig:GComparison}. Three choices of $G$ were chosen, all with the same $\langle x^2 \rangle$ and thus the same diffusion coefficient. The Gaussian, exponential, and power law commuter distribution all show an arrival time quadratic in position close to the origin with the same time offset. At later times, the Gaussian and exponential commuter distributions give rise to waves approximated by \eqref{eq:speed}, but the power law does not, consistent with \eqref{eq:Mollison}. The diffusion approximation seems to hold at small distances even for the power law with $p=4$ (in this case $\langle x^2 \rangle$ is finite), but because the higher-order moments diverge for the power law, the diffusion solution fails at larger distances. This rapid spread of infection for delocalized distributions is characterized by a sublinear arrival time. The fact that the offset time is the same for these three commuting distributions despite the power law behaving differently suggests that not all of the spreading dynamics can be captured by the offset time alone. One can compare the results from the power law to a completely delocalized contact distribution $G=1/L$ where the arrival time is constant and given by the time offset \eqref{eq:tOffsetWM}.

Despite the variety of solutions predicted by different choices of $G$, we expect infectious spread to be no faster than the well-mixed results and no slower than the diffusion results at a comparable $\langle x^2 \rangle$, when it is defined. This idea is explored further in figure \ref{fig:ArrivalTime(r0)}a where several different power law commuting distributions with the same $\sqrt{\langle x^2 \rangle}$ give the same offset time and early time dynamics, but far from the origin, the larger values of $p$ have larger arrival times.  For $p\leq 3$ in 1D, $\sqrt{\langle x^2 \rangle}$ is undefined, but we can instead fix the length scale $r_0$ in equation \eqref{eq:TruncatedPowerLaw} (with $l\rightarrow \infty$). See figure \ref{fig:ArrivalTime(r0)}b. For $p\leq 1$, the commuting distribution is not normalizable in 1D, so the dynamics are homogeneous. For $1<p\leq 3$, all moments diverge, and the offset time is noticeably larger while the arrival times at long distances are small. Larger values of $p$ have a smaller time offset, but arrive later at large $x$.

\begin{figure*}
    \centering
    \includegraphics[width=.75\textwidth]{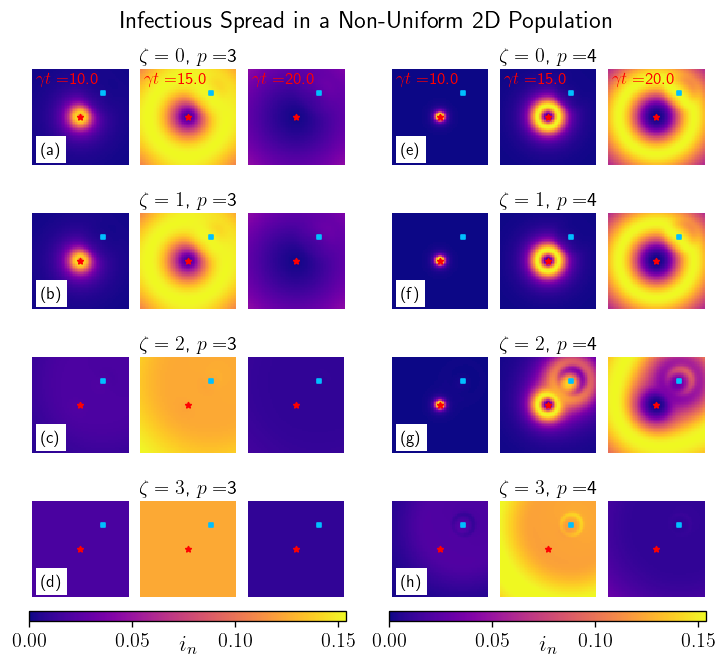}
    \caption{Simulations using a commuting probability $P_{nm} \sim N_m^\zeta(|\vec{x}_n-\vec{x}_m|+r_0)^{-p}$ for different choices of $\zeta$ and $p$ with $N_m = 100+9900\exp{-|\vec{x}_m-(.5,.5)|^2/\lambda_N^2}$ such that there is a city located in the northeast (blue square) with population 100 times larger than elsewhere in the domain. A single infected person is introduced at the origin (red star) at $t=0$. For each choice of $\zeta$ and $p$, three time snapshots of the infection fraction are shown. Panel (g) demonstrates a new phenomenon where the infection fraction first increases in the city then spreads away from the city. This behavior is not predicted in the diffusion approximation or the well-mixed approximation. The lattice is a 61 by 61 grid with side length $L=2$, so $dx=2/61$. Other parameters are $R_0=2$, $r_0=dx/4$, $\lambda_N=.05$.}
    \label{fig:PowerLaw}
\end{figure*}

To investigate the role of $\zeta$ for delocalized distributions, we will look at a two-dimensional nonuniform population and try different power laws and values of $\zeta$.  Numerical results are shown in figure \ref{fig:PowerLaw}. For $p \leq 4 $, $\langle x^2 \rangle$ diverges in two dimensions, so the arrival time grows slower than $x$. For $p=3$, some spatial dependence can be seen for small values of $\zeta$, but the dynamics are still nearly homogeneous. A symmetry appears to exist between the $\zeta=0$ and $\zeta=1$ cases for all values of $p$ studied, as we saw for the localized case. The case $\zeta=2$ and $p=4$ is the most interesting. The infection fraction increases along the line connecting the origin to the city, but grows slowest on the side of the city opposite the origin, producing a visible eclipse in the infection fraction. This can roughly be understood as a competition between two effects. $I(x,t)$ drifts toward the city from the origin as more people travel to the city to go to work than elsewhere, but there is the additional factor of $1/N(x)$ which reduces $i(x,t)$ close to the city. Once the infection reaches the city, there is a rapid spread from the city to the boundary. 

For large enough values of $\zeta$ for any $p$, the spreading appears homogeneous. To understand this, note that if $N_m$ has a unique maximum at $m=M$ and $G(|\vec{x}_m-\vec{x}_n|)$ is everywhere nonzero, then
\begin{equation}
    \lim_{\zeta \rightarrow \infty} P_{nm} = \delta_{mM} \implies \lim_{\zeta \rightarrow \infty} \sum_k \beta_{nk} i_k = \beta \frac{I_{\text{tot}}}{N_{\text{tot}}}. \label{eq:zetaInfinity}
\end{equation}
In words, for large $\zeta$, the entire population meets at the high-population hub where it is equally likely for someone to meet anyone else in the population, so the well-mixed equations \eqref{eq:SIR} are recovered. Typically we think of a well-mixed population as one where individuals travel everywhere, yet here, we see the emergent dynamics are the same if all commuters travel to one central location. In this limit, most interactions are second-order, so we will not see this effect using the simplified commuter interactions \eqref{eq:PR(IC)simplified} nor would this limit make sense in an Eulerian model. Although taking the limit $\zeta \rightarrow \infty$ seems unphysical, figure \ref{fig:PowerLaw} suggests that we don't need to make $\zeta$ too large to observe this effect. 

Little is known about epidemic models with delocalized contact distributions. Brockmann suggests that these cases can be modeled with fractional diffusion equations \cite{Brockmann_2010}. Some effort has been made to characterize the front dynamics in models with growth and fractional diffusion \cite{del-Castillo-Negrete_Carreras_Lynch_2003, Hanert_Schumacher_Deleersnijder_2011}, but little effort has been made to characterize the subtle arrival time properties we have seen here. Hallatschek and Fisher studied the arrival times in a stochastic model with delocalized dispersal patterns and found accelerating fronts \cite{Hallatschek_Fisher_2014}. They found distinct scaling regimes for the arrival time according to the power law of the hopping rate. Our model is distinct from these models in that it incorporates Lagrangian movement patterns. A detailed analysis of delocalized commuting in an epidemic model (to the level that we were able to do for the localized commuter distributions) is still an open problem.

\section{Discrete results}
Now that we understand the commuter model in the continuum, we would like to apply the model on a map using real geographic and census data. First, we will demonstrate new subtleties that arise when relaxing the assumptions of infinitesimal county size and uniform county shape. Then, the theory will be applied to a couple realistic commuting distributions. 

\begin{figure*}
    \centering
    \includegraphics[width=.85\textwidth]{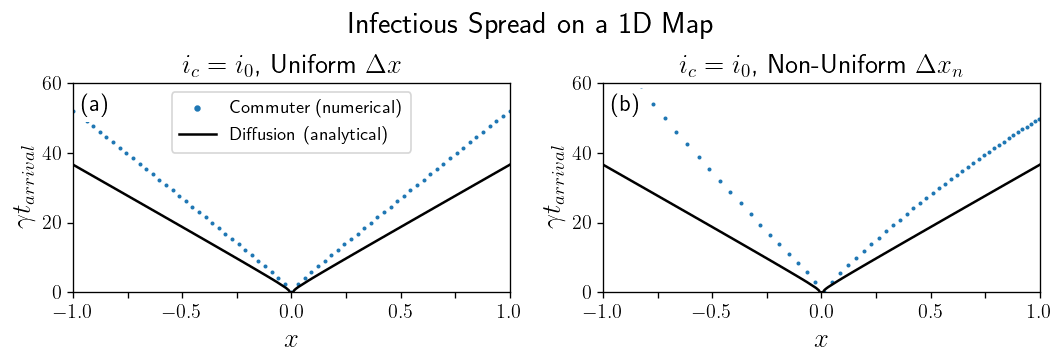}
    \caption{(a) Arrival times in $d=1$ using the same parameters as in figure \ref{fig:1DInfectiousSpread}, except $\Delta x=.03 > l$ such that the continuum theory (black line) fails. (b) The county spacing $\Delta x_n$ is taken to be nonuniform: $\Delta x_n = (\Delta x)e^{-4x_n/3L}$ such that the continuum theory fails and the infection reaches smaller counties earlier.}
    \label{fig:1Ddiscrete}
\end{figure*}

\subsection{Commuters on a non-uniform grid}
Let's work in the limit where the discreteness of the lattice is apparent. When the step size is larger than the characteristic commuting length, the diagonal entries of $\mathbf{P}$ dominate, so the infection spreads slower than the results obtained by diffusion. See figure \ref{fig:1Ddiscrete}a. The offset time is reduced because a larger step size means the growth term dominates the diffusion term early on. Note that the value of $\langle x^2 \rangle$ used for the analytical result is that obtained from the continuous function $G(|\vec{x}|)$. Infection waves are still observed at late times, but with a speed less than \eqref{eq:speed}. 

An additional complication which arises on maps is that counties may all be different sizes. Because the distance to the center of a neighboring county with smaller area will be less than the distance to the center of a neighboring county with larger area, the probability to travel to a smaller county will be greater. This effect is demonstrated in figure \ref{fig:1Ddiscrete}b where a non-uniform county spacing causes the disease to drift rightward.

These complications highlight the care that one must take when applying the model to maps. Unlike some biological systems where discreteness is a relevant feature that leads to interesting spreading phenomena \cite{Dieterle_Amir_2021}, the spread of disease should not depend on county borders, so these effects must be artificial. The problem arises because the commuting distribution is defined in continuous space $G(|\vec{y}-\Vec{x}|)$, but the populations ($N_n, I_n, $ etc.) are defined for discrete systems. In order to construct the commuting probabilities between counties, we assumed the form \eqref{eq:mobility} where $\Vec{x}_n, \vec{x}_m$ are the county centers. This is a good estimate when $\Delta x$ is small, but in cases where $\Delta x$ is large, our expressions should involve integrals over the county area. There are two solutions. Of course, if the commuting probabilities or commuting fluxes between counties are known, then guessing a form such as \eqref{eq:mobility} is unnecessary, and solving the discrete model \eqref{eq:CommuterModelSIR} is appropriate. For the purposes of our theory where we are investigating the role of different gravity models, one can coarse grain the populations so that the continuum approximation holds, find the density fields using \eqref{eq:iC}, then integrate the fields over the county area to obtain $S_n, I_n, R_n$. In the next section, when we apply the model to the United States, we find the qualitative results of our continuum theory to hold when naively applying \eqref{eq:mobility}, so at the cost of not getting quantitative agreement with our analytic results, we will use the simpler approach of evaluating the commuting probability at the center of population of each county.

\subsection{Commuters on a map}
We simulate the commuter model on a map of the United States with realistic populations and county geometries. Population data was obtained from the U.S. Census Bureau \cite{Census}. Geometric data was retrieved from Plotly \cite{plotly} and visualized using the GeoPandas package of Python. The coordinate $\vec{x}_n$ of a given county is taken to be the center of population. Given the latitudes $\{\theta_n \}$ and longitudes $\{\phi_n\}$ of each county center in radians, the separation between two counties is approximated as 
\begin{equation}
    |\vec{x}_m-\vec{x}_n| \approx R_E \sqrt{\left(\theta_m-\theta_n\right)^2 + \cos^2 \left(\frac{\theta_m + \theta_n}{2} \right)(\phi_m - \phi_n)^2}.
\end{equation}

\begin{figure*}
    \centering
    \includegraphics[width=.9\textwidth]{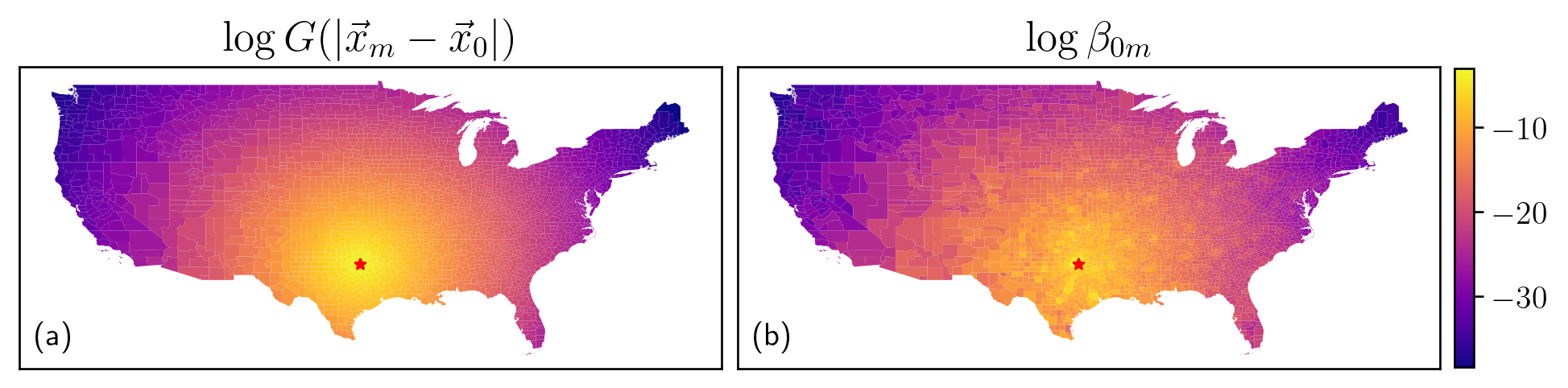}
    \includegraphics[width=.9\textwidth]{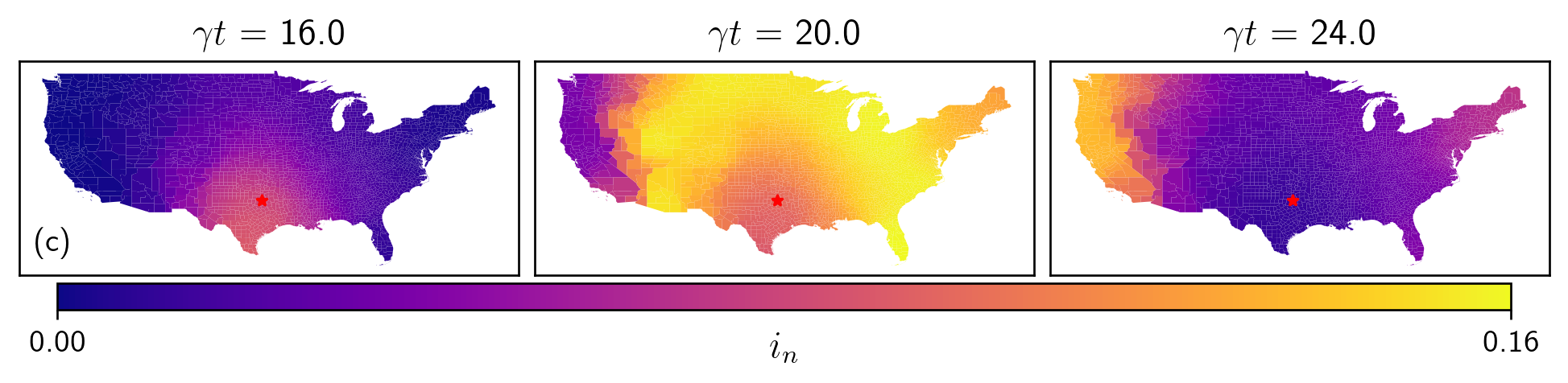}
    \includegraphics[width=.9\textwidth]{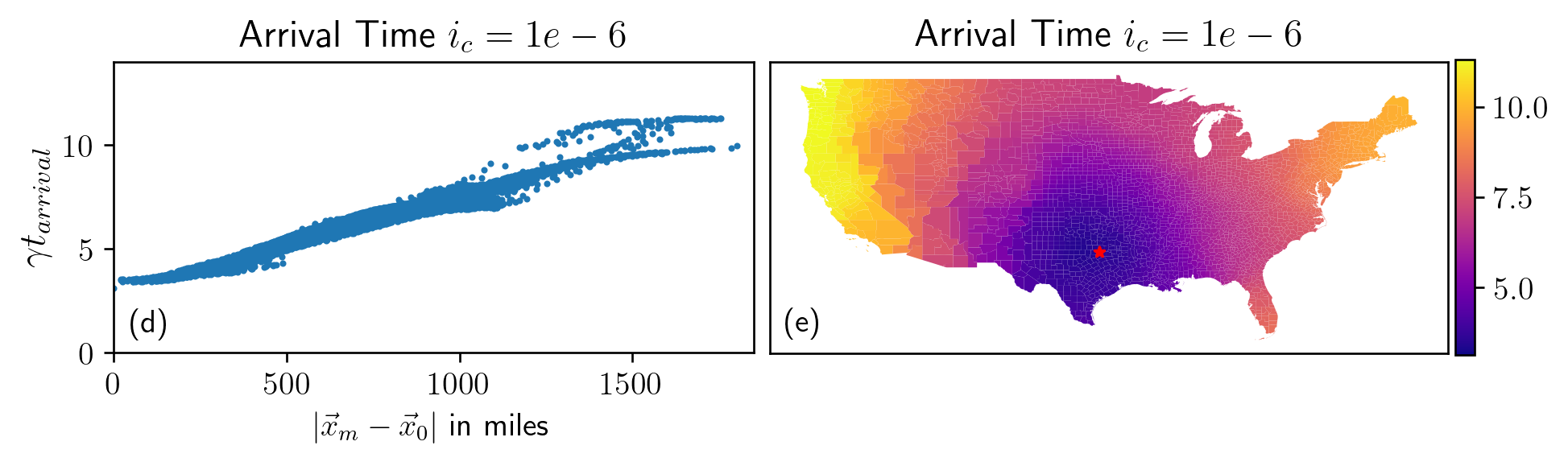}
    \caption{(a)-(b) Commuting distribution and derived contact matrix elements measured with respect to Dallas, TX (star) using the the commuting parameters obtained from \cite{Balcan_Colizza_Goncalves_Hu_Ramasco_Vespignani_2009}: $P_{nm} \sim N_m^\zeta e^{-|\vec{x}_n-\vec{x}_m|/l}$, $l=51$ mi, $\zeta = .64$. (c) Infectious spread from Dallas, TX with $R_0=2$ and $I_0=1$. Three time snapshots of the infection fraction are shown. (d) Arrival time as a function of distance from Dallas and (e) displayed on the map.}
    \label{fig:colizza}
\end{figure*}

The first set of parameters we will try is that obtained by Balcan, et al. \cite{Balcan_Colizza_Goncalves_Hu_Ramasco_Vespignani_2009} who performed a Voronoi tessellation based on distance from major airports to divide the United States into units more suitable for studying mobility processes. They found that the commuting distribution takes the form of an exponential with a decay length of $l=51$ miles (for distances not too large). We will try applying this commuting distribution at the county level, though it's worth noting that because the spatial units considered in their paper were larger than a typical U.S. county, this may be an unrealistically large $l$ for our purposes. The commuting distribution and resulting contact matrix elements are visualized in figures \ref{fig:colizza}a and \ref{fig:colizza}b. Broad infection waves are shown in figure \ref{fig:colizza}c. Even though $l$ is large, figure \ref{fig:colizza}d shows that the arrival time still resembles that predicted by the reaction-diffusion system: a finite offset followed by a quadratic regime and finally a linear regime. Also, notice that even though the commuting probability is asymmetric due to the non-uniform population distribution in the U.S., the arrival time changes smoothly, as observed by the smooth gradient in the colors of figure \ref{fig:colizza}e.

\begin{figure*}
    \centering
    \includegraphics[width=.9\textwidth]{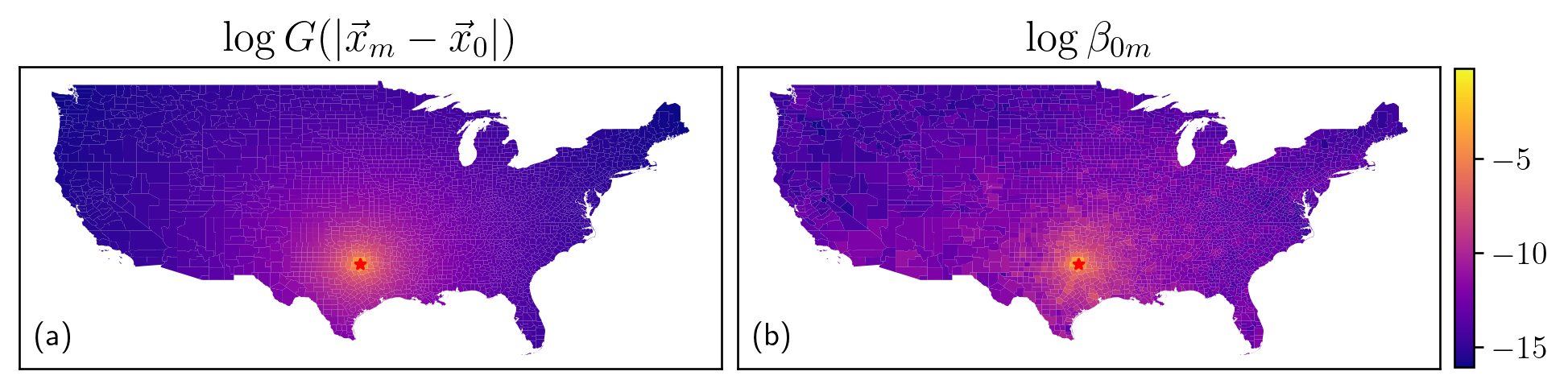}
    \includegraphics[width=.9\textwidth]{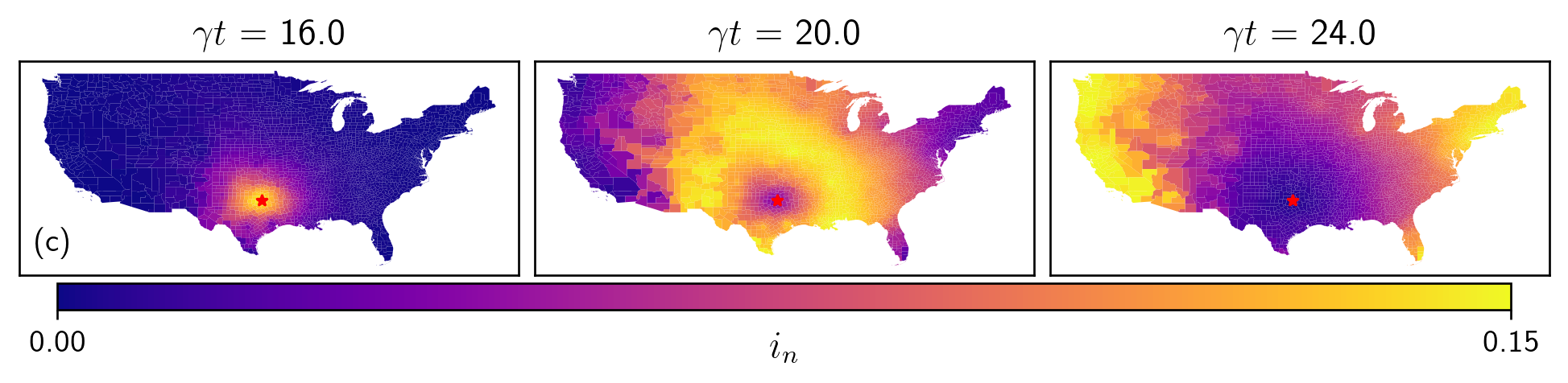}
    \includegraphics[width=.9\textwidth]{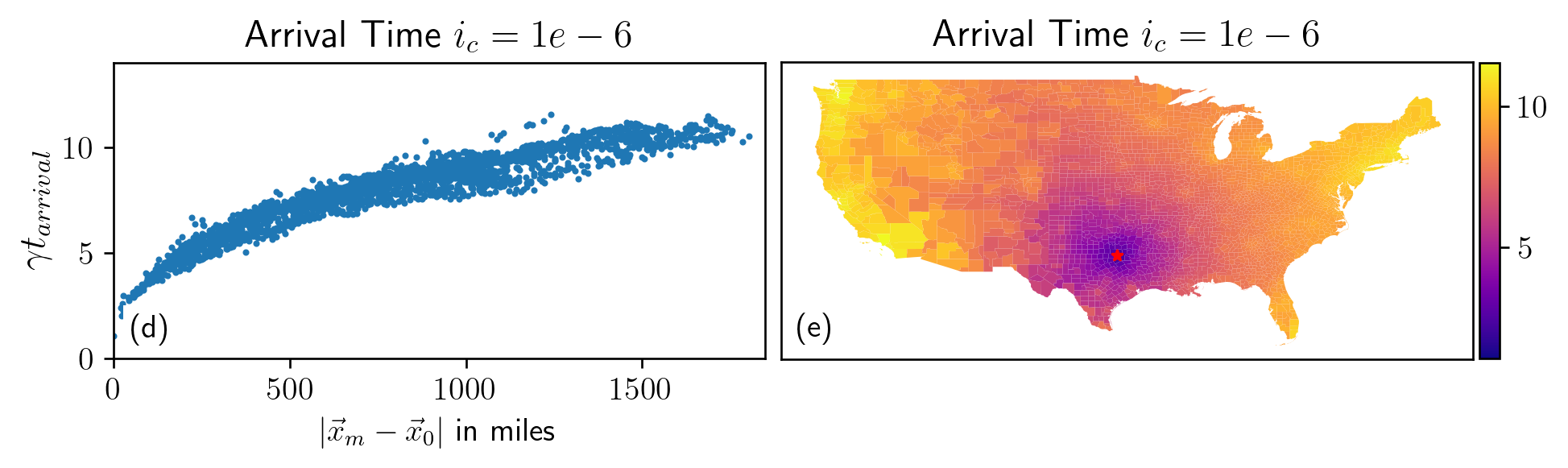}
    \caption{(a)-(b) Commuting distribution and derived contact matrix elements measured with respect to Dallas, TX (star) using the the commuting parameters obtained from \cite{Viboud_Bjornstad_Smith_Simonsen_Miller_Grenfell_2006}: $P_{nm} \sim N_m^\zeta (|\vec{x}_n-\vec{x}_m|+r_0)^{-p}$, $p=3.05$, $\zeta = .64$. The small-scale cutoff (not considered in their paper) was chosen to be $r_0=10$ mi. (c) Infectious spread from Dallas, TX with $R_0=2$ and $I_0=1$. Three time snapshots of the infection fraction are shown. (d) Arrival time as a function of distance from Dallas and (e) displayed on the map.}
    \label{fig:viboud}
\end{figure*}

Now, we will consider the parameters obtained from county workflow data by Viboud, et al. \cite{Viboud_Bjornstad_Smith_Simonsen_Miller_Grenfell_2006}. They found a power law fit for the commuting distribution with $p=3.05$, and the tendency to travel to cities is given by the exponent $\zeta=.64$ (which is, interestingly, identical to the value that Balcan found \cite{Balcan_Colizza_Goncalves_Hu_Ramasco_Vespignani_2009}). Results are shown in figure \ref{fig:viboud}. The arrival time is sublinear and has a larger variance. As expected with delocalized commuting distributions, the disease travels faster than any wave, and complex population-dependent behavior emerges. 

\begin{figure*}
    \centering
    \includegraphics[width=.9\textwidth]{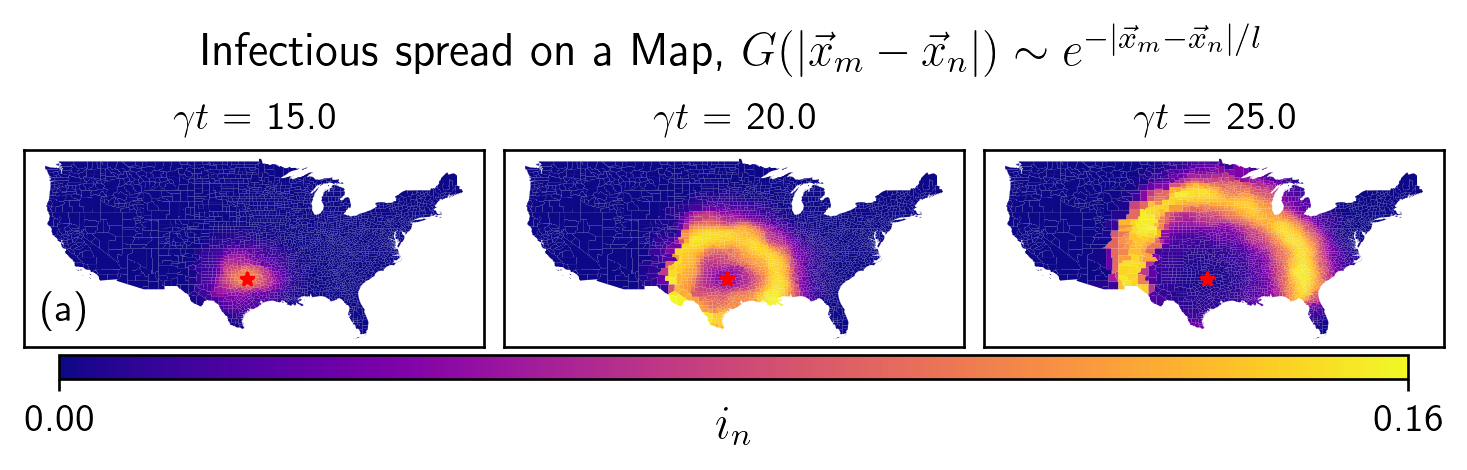}
    \includegraphics[width=.9\textwidth]{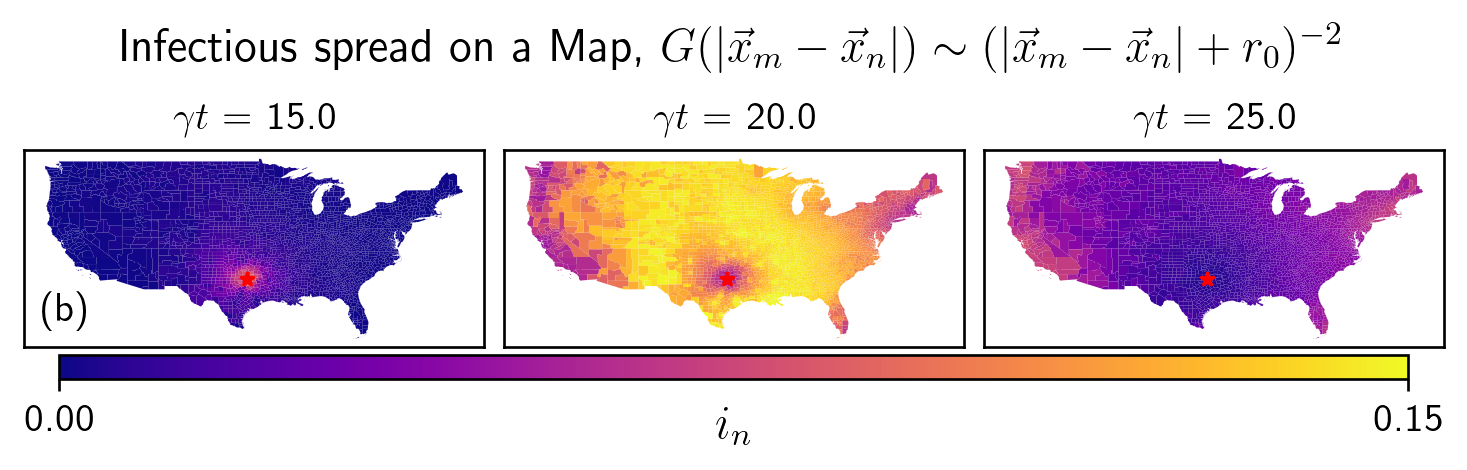}
    \caption{(a) Infectious spread from Dallas, TX (star) for a localized commuting distribution with $l=15$ miles. (b) Infectious spread from Dallas, TX for a delocalized ($p=2$) commuting distribution with $r_0=1$ mile. Additional parameters used for these simulations were $\zeta=0$, $R_0=2$, and $I_0=1$. Three time snapshots of the infection fraction are shown for each case. The localized commuting distribution produces traveling waves from the infection source while the delocalized commuting distribution leads to nearly homogeneous infection growth across the country.}
    \label{fig:mapDemo}
\end{figure*}

These two examples demonstrate that although some human mobility processes can seemingly be described by either an exponential distribution or a power law, the emergent infection dynamics will differ greatly. This claim is even stronger when comparing commuter distributions with smaller length scales, as shown in figure \ref{fig:mapDemo}. At $\gamma t=15$, the epidemic appears more severe close to Dallas for the localized distribution (figure \ref{fig:mapDemo}a) due to the smaller time offset, but by $\gamma t=20$, the infection has spread across the entire country for the delocalized distribution (figure \ref{fig:mapDemo}b), while the infection spreads as a wave for the localized distribution. In either case, the final fraction of individuals who were at some point infected was found to be $r_{\text{tot}}=.797$, identical to that predicted by \eqref{eq:rinf}, as expected.

\section{Discussion}

Even in our globalized world, commuting patterns play an important role in shaping disease spread. During the beginning of the COVID-19 pandemic in early 2020, commuting flows helped to explain the spread of disease from its source in Wuhan, China \cite{Jia_Lu_Yuan_Xu_Jia_Christakis_2020}. Spatiotemporal data in Italy suggests that the infection traveled along highways \cite{Gatto_Bertuzzo_Mari_Miccoli_Carraro_Casagrandi_Rinaldo_2020}. Recent work showed that metapopulation models can be used to quantitatively describe the spread of SARS-CoV-2 \cite{Chang_Pierson_Koh_Gerardin_Redbird_Grusky_Leskovec_2021, Treut_Huber_Kamb_Kawagoe_McGeever_Miller_Pnini_Veytsman_Yllanes_2021}. These metapopulation models are closely related to the classic papers on disease propagation \cite{Kendall_1965,Mollison_1972a, Mollison_1972b}, but possess new challenges. Little work has been done to bridge the gap between the more data-driven metapopulation models and the theoretical results derived in mathematical epidemiology. One work made this connection in the Eulerian formalism \cite{Brockmann_2010}, less appropriate for incorporating commuter dynamics. These Eulerian movement models have been more useful for deriving effective metrics used to characterize the arrival times of a disease on a network \cite{Brockmann_Helbing_2013, Iannelli_Koher_Brockmann_Hovel_Sokolov_2017}, but infection dispersal due to commutes gives rise to behaviors that are best studied in real space using a Lagrangian movement model where individuals have a well-defined home. While Lagrangian movement for commuters has been incorporated into epidemic models \cite{Calvetti_Hoover_Rose_Somersalo_2021, Charaudeau_Pakdaman_Boelle_2014, Treut_Huber_Kamb_Kawagoe_McGeever_Miller_Pnini_Veytsman_Yllanes_2021}, these works focus on contact matrices derived from known commuting patterns of specific regions. Here, we studied a broad class of models that incorporate more general principles from human mobility theory \cite{Barbosa_2018,Gallotti_Bazzani_Rambaldi_Barthelemy_2016}.

Two commonly studied limits were obtained from our model. When the commuting distribution has an exponential tail and the domain is large enough, the infection is initially localized near the source and the spread is diffusive, but at later times, traveling infection waves can be observed. Perhaps surprisingly, the infection fraction wave drifts toward lower populations, so an individual in a small town is initially at a higher risk of getting sick than an individual in a big city, even if the number of cases is higher in the city. If the infection quickly reaches the boundary (which could be the result of a long-tailed commuting distribution or an infection wave front whose size is comparable to the size of the domain), then the disease will be undetected for a characteristic offset time, but once it is detected at one location, it will soon be detected everywhere. The former case can be modeled using a system of reaction-diffusion equations, and the latter using the spatially independent $SIR$ model. In these two limits, the exact form of the commuting probabilities is unimportant. However, there is an intermediate regime where both the exact commuting distribution and the asymmetry in the commuting probability matter. In order to correctly capture this intermediate regime, we had to consider the commuting dynamics of both the susceptible and the infected individuals, leading to a contact matrix that is nonlinear in the commuting probability. 

Although we have not focused on the spreading due to long-distance air travel, such dynamics could be incorporated into the commuting probabilities. But, on scales where the dynamics are dominated by commuters, the spatially embedded infectious spread considered in this paper will dominate. While the mathematical literature tends to focus on the limiting behavior far from the infection source, realistically, the infection may never achieve this limit on scales where the commuter dynamics dominate. For that reason, our paper has focused on both the short-term and long-term arrival time behavior. In particular, the offset time is defined for even a single county, so the techniques developed here can elucidate the infectious spread at scales as small as the step size at which data is taken. The behavior close to the infection origin can be understood as a transition from an initial dispersal-dominated regime where the disease is spreading thin throughout counties near the origin to a growth-dominated regime where the infection at each location grows exponentially until it is detected. There is an interesting tradeoff in the short-term and long-term behavior of an epidemic. In cases where people travel far, the disease can go undetected for a longer period of time and case counts will be everywhere low early on, but once the disease is detected at one location, it will be detected everywhere and the entire population will be experiencing the epidemic simultaneously. If, however, people remain close to home, the disease will be detected sooner near the origin, but the infection will travel slower to neighboring counties, so the arrival time far from the origin is higher. In this way, the commuting patterns will dictate the times at which an infectious disease is detected at different locations when long-distance flights are neglected, but the total number of cases can only be lowered by reducing the number of contacts.

In characterizing the role of commuter distributions in disease spread, we made several simplifying assumptions that limit the predictive power of the commuter model. We assumed that the total number of contacts a person makes is independent of their home location and time, but one could easily relax this assumption. Our model is deterministic, so it neglects the possibility of finite-time extinction. We assumed zero latency period, lifetime immunity, and neglected births and deaths. The analysis presented here could be adapted to more complicated epidemic models or other social phenonomena such as rumor spreading \cite{Castellano_Fortunato_Loreto_2009}. While the precise dynamics will depend on any new rate constants introduced, the lessons persist: only localized commuting distributions give rise to local theories of dispersal, higher-order contact processes matter, and the early-time dynamics should not be ignored. 

\begin{acknowledgments}
This research was supported by NSF Grant No. PHY-1554887,  the  University  of  Pennsylvania  Materials Research Science and Engineering Center (MRSEC) through Grant No. DMR-1720530,
and the Simons Foundation through Grant No. 568888.\end{acknowledgments}

\appendix

\section{Derivation of the reaction-diffusion equation}
Beginning with \eqref{eq:iC}, we can derive the reaction-diffusion system \eqref{eq:reactionDiffusion} by performing a series of Taylor expansions. Crucially, this is only valid when \eqref{eq:Mollison} is satisfied, such that all of the moments of our commuting distribution converge. We first estimate the fraction of people currently located at $\vec{y}$ who are infected. Since $G(|\vec{z}-\vec{y}|)$ is exponentially localized near $\vec{z}=\vec{y}$, only the values of $i$ near $\vec{z}=\vec{y}$ will matter, so we can Taylor expand $i(\vec{z})$ around $i(\vec{y})$, and do the same for $N(\vec{z})$.

\begin{widetext}
\begin{align}
    \frac{\int d^d z \hspace{.1cm}P(\vec{z},\vec{y}) N(\vec{z}) i(\vec{z},t) }{\int d^d z \hspace{.1cm}P(\vec{z},\vec{y})  N(\vec{z})} &\approx i(\vec{y},t) + \frac{\int d^d z \hspace{.1cm}P(\vec{z},\vec{y}) N(\vec{z}) (z^\mu - y^\mu)}{\int d^d z \hspace{.1cm}P(\vec{z},\vec{y})  N(\vec{z})} \partial_\mu i (\vec{y},t) \nonumber \\&\hspace{1.3cm}+ \frac{1}{2} \frac{\int d^d z \hspace{.1cm}P(\vec{z},\vec{y}) N(\vec{z}) (z^\mu - y^\mu)(z^\nu - y^\nu)}{\int d^d z \hspace{.1cm}P(\vec{z},\vec{y})  N(\vec{z})} \partial_\mu \partial_\nu i (\vec{y},t) \nonumber \\
    &\approx i(\vec{y},t) + \int d^d z \hspace{.1cm}G(\vec{z},\vec{y}) \left[1+ (1-\zeta) \partial_\nu \log N(\vec{y}) (z^\nu - y^\nu)\right] (z^\mu - y^\mu) \partial_\mu i (\vec{y},t) \nonumber \\&\hspace{1.3cm}+ \frac{1}{2} \int d^d z \hspace{.1cm}G(\vec{z},\vec{y})  (z^\mu - y^\mu)(z^\nu - y^\nu) \partial_\mu \partial_\nu i (\vec{y},t) \nonumber\\
    &= i(\vec{y},t) + \frac{\langle y^2 \rangle}{d} (1-\zeta)  \grad \log N(\vec{y}) \cdot \grad i(\vec{y},t) + \frac{1}{2} \frac{\langle y^2 \rangle}{d}  \nabla^2 i(\vec{y},t) 
\end{align}
Using the same procedure, we can calculate the probability that a contact made by someone living at $\vec{x}$ is with an infected. 
\begin{align}
    \int d^d y \hspace{.1cm}P(\vec{x},\vec{y}) \frac{\int d^d z \hspace{.1cm}P(\vec{z},\vec{y}) N(\vec{z}) i(\vec{z},t) }{\int d^d z \hspace{.1cm}P(\vec{z},\vec{y})  N(z)} &\approx \int d^d y \hspace{.1cm}P(\vec{x},\vec{y})\nonumber \\
    &\hspace{.4cm}\cross \left[ i(\vec{y},t) + \frac{\langle y^2 \rangle}{d} (1-\zeta) \grad \log N(\vec{y}) \cdot \grad i(\vec{y},t) + \frac{1}{2} \frac{\langle y^2 \rangle}{d}  \nabla^2 i(\vec{y},t) \right] \nonumber \\
    &=\int d^d y \hspace{.1cm}G(\vec{x},\vec{y})[1+\zeta \partial_\nu \log N(\vec{x}) (y^\nu-x^\nu)]\nonumber \\& \hspace{.4cm}\cross\left[ i(\vec{x},t) + \partial_\mu i(\vec{x},t) (y^\mu-x^\mu)+\frac{1}{2}\partial_\mu \partial_\nu i(\vec{x},t)(y^\mu-x^\mu)(y^\nu-x^\nu)\right] \nonumber\\ &\hspace{.4cm}+ \frac{\langle x^2 \rangle}{d} (1-\zeta) \grad \log N(\vec{x}) \cdot \grad i(\vec{x},t) + \frac{1}{2} \frac{\langle x^2 \rangle}{d}  \nabla^2 i (\vec{x},t) \nonumber\\
    &= i(\vec{x},t) + \frac{\langle x^2 \rangle}{d} \grad \log N(\vec{x}) \cdot \grad i(\vec{x},t) + \frac{\langle x^2 \rangle}{d} \nabla^2 i(\vec{x},t) \label{eq:2ndOrderTaylorExpansion}
\end{align}
\end{widetext}
The same method was used to derive equation \eqref{eq:fourthorder} where terms up to fourth order in derivatives were kept.

\section{Table of parameters}
See table \ref{tab:parameters} for a summary of functions and parameters used in the paper. 

\begin{widetext}
    
\begin{table}
  \begin{center}
\def~{\hphantom{0}}

\begin{tabular}{c  c} 
 \hline
 Symbol & Description \\
 \hline\hline
 $S_n(t), I_n(t), R_n(t)$ & (S)usceptible, (I)nfected, and (R)ecovered population in county $n$\\ 
 $N_n$ & Population of county $n$ \\ 
  $s_n(t), i_n(t), r_n(t)$ & $S_n(t)/N_n, I_n(t)/N_n, R_n(t)/N_n$ \\ 
   $S(\Vec{x},t),I(\Vec{x},t),R(\Vec{x},t)$ & (S)usceptible, (I)nfected, and (R)ecovered population density at $\vec{x}$ in continuous space\\ 
   $N(\Vec{x})$ & Population density at $\Vec{x}$ in continuous space \\
   $s(\Vec{x},t),i(\Vec{x},t),r(\Vec{x},t)$ & $S(\Vec{x},t)/N(\Vec{x}),I(\Vec{x},t)/N(\Vec{x}),R(\Vec{x},t)/N(\Vec{x})$ \\ $S_{\text{tot}}(t), I_{\text{tot}}(t), R_{\text{tot}}(t)$ & $\sum_n S_n(t),\sum_n I_n(t),\sum_n R_n(t)$\\
   $N_{\text{tot}}$ & $\sum_n N_n$ \\
    $s_{\text{tot}}(t), i_{\text{tot}}(t), r_{\text{tot}}(t)$ & $S_{\text{tot}}(t)/N_{\text{tot}}, I_{\text{tot}}(t)/N_{\text{tot}}, R_{\text{tot}}(t)/N_{\text{tot}} $\\
   $I_0$ & Initial infected population: $I(\vec{x},0) = I_0 \delta(\vec{x},0)$ or $I_n(0) = I_0 \delta_{n0}$ \\
   $G(|\vec{x}|)$ & Commuting distribution, often taking the form $G(|\vec{x}|) \sim (|\vec{x}|+r_0)^{-p} e^{-|\vec{x}|/l}$\\
   $P_{nm}$ & Commuting probability, often taking the form of a gravity law $P_{nm}\sim N_m^\zeta G(|\Vec{x}_m-\Vec{x}_n|)$\\
   $l, r_0, p, \zeta$ & Commuting parameters defined by the above two equations\\
    $\beta_{nk}$ & Contact matrix related to $P_{nm}$ by equation \eqref{eq:contactMtxPnm}\\ 
    $\beta$ & Contact rate $\sum_k \beta_{nk}$, assumed constant\\ 
    $\gamma$ & Recovery rate \\ 
    $R_0$ & Basic reproduction number $\beta/\gamma$ \\ 
     $\langle x^2 \rangle$ & Second moment of $G$ \\ 
    $\Delta x$ & Characteristic county size\\ 
    $L$ & System size\\
    $d$ &  Spatial dimension \\ 
    $t_{\text{arrival}}(\vec{x}_n)$ &  Arrival time at $\vec{x}_n$ satisfying $I_n(t_{\text{arrival}}) = I_c$, or $i_n(t_{\text{arrival}}) = i_c$\\     
    $t_O$ &  Offset time, the smallest time satisfying $I(0,t_O)\geq I_c$ and $\dot{I}(0,t_O) \geq 0$ \\     
    $D$ & Diffusion coefficient given by $\beta \langle x^2 \rangle/d$ \\
    $\vec{v}$ & Drift velocity given by $D \grad \log N$\\
    $c$ &  Wave speed obtained by minimizing \eqref{eq:speedCorrected}, approximately given by $\sqrt{4D(\beta - \gamma)}$ \\
    \hline \hline
 
\end{tabular}
  \caption{Parameters used in the paper.}
  \label{tab:parameters}
  \end{center}
\end{table}

\end{widetext}

\bibliography{refs}

\end{document}